\documentclass[AMS,STIX1COL]{WileyNJD-v2}

\articletype{Survey Article}%

\received{31 November 2022}
\revised{XX YY 2023}
\accepted{XX YY 2023}

\usepackage[utf8]{inputenc}
\usepackage[a-1b]{pdfx}   
\usepackage{graphicx}
\usepackage{subcaption}
\usepackage{times}
\usepackage{epsfig}
\usepackage{color}
\usepackage{xcolor}
\usepackage{graphicx}
\usepackage{wrapfig}
\usepackage{amssymb, amsmath}
\usepackage{multirow}
\usepackage{framed}
\usepackage{url}
\usepackage{listings}
\usepackage{booktabs} 
\usepackage{enumitem}
\usepackage{comment}
\usepackage{graphicx}
\graphicspath{ {./Images/} }

\newcommand{\afonso}[1]{{\color{red}\sc Afonso: #1}}
\usepackage{algorithm}
\usepackage{longtable}
\usepackage{tabularx}
\usepackage{xspace}

\def\papers{124\xspace} 
\def\papersinitial{119\xspace} 
\def\paperssnowballing{five\xspace} 
\def\papersinput{82\xspace}
\def\papersoracle{42\xspace}

\def\paperssystem{33\xspace} 
\def\paperssystemblack{25\xspace}
\def\paperssystemwhite{8\xspace}
\def\papersgui{24\xspace} 
\def\papersunit{11\xspace} 
\def\papersunitblack{2\xspace}
\def\papersunitwhite{9\xspace}
\def\papersperformance{9\xspace} 
\def\paperscit{5\xspace} 

\def\papersverdict{12\xspace} 
\def\papersexpected{20\xspace} 
\def\papersmetamorphic{10\xspace} 

\def\papersblackbox{65\xspace} 
\def\paperswhitebox{17\xspace} 

\begin{document}

\title{The Integration of Machine Learning into Automated Test Generation: A Systematic Mapping Study}

\author[1]{Afonso Fontes, Gregory Gay*}

\authormark{Fontes and Gay}

\address[1]{\orgdiv{Department of Computer Science and Engineering}, \orgname{Chalmers $\vert$ University of Gothenburg}, \orgaddress{\country{Sweden}}}

\corres{*Gregory Gay, \email{greg@greggay.com}}

\abstract[Abstract]{
\noindent\textbf{Context:} Machine learning (ML) may enable effective automated test generation.

\noindent\textbf{Objectives:} We characterize emerging research, examining testing practices, researcher goals, ML techniques applied, evaluation, and challenges.

\noindent\textbf{Methods:} We perform a systematic mapping study on a sample of \papers publications.

\noindent\textbf{Results:} ML generates input for system, GUI, unit, performance, and combinatorial testing or improves the performance of existing generation methods. ML is also used to generate test verdicts, property-based, and expected output oracles. Supervised learning---often based on neural networks---and reinforcement learning---often based on Q-learning---are common, and some publications also employ unsupervised or semi-supervised learning. (Semi-/Un-)Supervised approaches are evaluated using both traditional testing metrics and ML-related metrics (e.g., accuracy), while reinforcement learning is often evaluated using testing metrics tied to the reward function. 

\noindent\textbf{Conclusion:} Work-to-date shows great promise, but there are open challenges regarding training data, retraining, scalability, evaluation complexity, ML algorithms employed---and how they are applied---benchmarks, and replicability. Our findings can serve as a roadmap and inspiration for researchers in this field.
}


\maketitle

\section{Introduction}\label{sec:intro}

\textit{Software testing} is invaluable in ensuring the reliability of the software that powers our society~\cite{testOracleSurvey2014}. It is also notoriously difficult and expensive, with severe consequences for productivity, the environment, and human life if not conducted properly. New tools and methodologies are needed to control that cost without reducing the quality of the testing process.  

Automation has a critical role in controlling costs and focusing developer attention~\cite{Orso14:STR}. Consider test generation---an effort-intensive task where sequences of program \textit{input} and \textit{oracles} that judge the correctness of the resulting execution are crafted for a system-under-test (SUT)~\cite{testOracleSurvey2014}. Effective automated test generation could lead to immense effort and cost savings.

Automated test generation is a popular research topic, and outstanding achievements have been made in the area~\cite{Orso14:STR}. Still, there are critical limitations to current approaches. Major among these is that generation frameworks are applied in a \textit{general} manner---techniques target simple universal heuristics, and those heuristics are applied in a static manner to all systems equally. Parameters of test generation can be tuned by a developer, but this requires advanced knowledge and is still based on the same universal heuristics. Current generation frameworks are largely unable to adapt their approach to a particular SUT, even though such projects offer rich information content in their documentation, metadata, source code, or execution logs~\cite{surveyMLinTesting2019}. Such static application limits the potential effectiveness of automated test generation. 

Machine learning (ML) algorithms make predictions by analyzing and extrapolating from sets of observations~\cite{surveyMLinTesting2019}. Advances in ML have shown that automation can match or surpass human performance across many problem domains. ML has advanced the state-of-the-art in virtually every field. Automated test generation is no exception. Recently, researchers have begun to use ML either to \textit{directly} generate input or oracles~\cite{Kim2018} or to \textit{enhance} the effectiveness or efficiency of existing test generation frameworks~\cite{Almulla2020}. ML offers the potential means to adapt test generation to a SUT, and to enable automation to optimize its approach without human intervention. 

We are interested in understanding and characterizing emerging research around the integration of ML into automated test generation\footnote{We focus specifically on the use of ML to enhance test generation, as part of the broader field of AI-for-Software Engineering (AI4SE). There has also been research in automated test generation for ML-based systems (SE4AI). These studies are out of the scope of our review.}. Specifically, we are interested in which testing practices have been addressed by integrating ML into test generation, the goals of the researchers using ML, how ML is integrated into the generation process, which specific ML techniques are applied, how such techniques are trained and validated, and how the whole test generation process is evaluated. We are also interested in identifying the emerging field's limitations and open research challenges. To that end, we have performed a systematic mapping study. Following a search of relevant databases and a rigorous filtering process, we have examined \papers relevant studies, gathering the data needed to answer our research questions.

We observed that ML supports generation of input and oracles for a variety of testing practices (e.g., system or GUI testing) and oracle types (e.g., expected test verdicts and expected output values). During input generation, ML either directly generates input or improves the efficiency or effectiveness of existing generation methods. The most common types of ML are supervised and reinforcement learning. A small number of publications also employ unsupervised or semi-supervised/adversarial learning. 

Supervised learning is the most common type for system testing, Combinatorial Interaction Testing, and all forms of oracle generation. Neural networks are the most common supervised techniques, and techniques are evaluated using both traditional testing metrics (e.g., coverage) and ML metrics (e.g., accuracy). Reinforcement learning is the most common ML type for GUI, unit, and performance testing. It is effective for practices with scoring functions and when testing requires a sequence of input steps. It is also effective at tuning generation tools. Reinforcement learning techniques are generally based on Q-Learning, and are generally evaluated using testing metrics tied to the reward function. Finally, unsupervised learning is effective for filtering tasks such as discarding similar test cases. 

The publications show great promise, but there are significant open challenges. Learning is limited by the required quantity, quality, and contents of training data. Models should be retrained over time. Whether techniques will scale to real-world systems is not clear. Researchers rarely justify the choice of ML technique or compare alternatives. Research is limited by the overuse of simplistic examples, the lack of standard benchmarks, and the unavailability of code and data. Researchers should be encouraged to use common benchmarks and provide replication packages and code. In addition, new benchmarks could be created for ML challenges (e.g., oracle generation). 

Our study is the first to thoroughly summarize and characterize this emerging research field\footnote{This publication extends an initial systematic literature review~\cite{Fontes21:SLR} that focused only on test oracle generation. Our extended study also includes publications on test input generation and an expanded set of publications for oracle generation. We also include additional and extended analyses and discussion.} We hope that our findings will serve as a roadmap for both researchers and practitioners interested in the use of ML in test generation and that it will inspire new advances in the field.

\section{Background and Related Work}\label{sec:background}


\subsection{Software Testing}\label{sec:test_intro}

It is essential to verify that software functions as intended. This verification process usually involves \textit{testing}---the application of \textit{input}, and analysis of the resulting \textit{output}, to identify unexpected behaviors in the system-under-test (SUT)~\cite{testOracleSurvey2014}.

During testing, a \textit{test suite} containing one or more \textit{test cases} is applied to the SUT. A test case consists of a \textit{test sequence (or procedure)}--a series of interactions with the SUT--with \textit{test input} applied to some SUT component. Depending on the granularity of testing, the input can range from method calls, to API calls, to actions within a graphical interface. Then, the test case will validate the output against a set of encoded expectations---the \textit{test oracle}---to determine whether the test passes or fails~\cite{testOracleSurvey2014}. An oracle can be a predefined specification (e.g., an assertion), output from a past version, a model, or even manual inspection by humans~\cite{testOracleSurvey2014}. 

\begin{figure}[!t]
	\centering
	\begin{lstlisting}[language=Java,basicstyle={\scriptsize\ttfamily}]
	@Test
	public void testPrintMessage() {
	    String str = "Test Message";
	    TransformCase tCase = new TransformCase(str);
	    String upperCaseStr = str.toUpperCase();
	    assertEquals(upperCaseStr, tCase.getText());
	}
	\end{lstlisting}
	\caption{Example of a unit test case written using the JUnit notation for Java.}
	\label{fig:testcase}
\end{figure}


An example of a test case, written in the JUnit notation, is shown in Figure~\ref{fig:testcase}. The test input is a string passed to the constructor of the \texttt{TransformCase} class, then a call to \texttt{getText()}. An assertion then checks whether the output matches the expected output---an upper-case version of the input. 

Testing can be performed at different granularity levels, using tests written in code or applied by humans. 
The lowest granularity is unit testing, which focuses on isolated code modules (generally classes). Module interactions are tested during integration testing. Then, during system testing, the SUT is tested through one of its defined interfaces---a programmable interface, a command-line interface, a graphical user interface, or another external interface. Human-driven testing, such as exploratory testing, is out of the scope of this study, as it is often not amenable to automation.

\subsection{Machine Learning}\label{sec:MLintro}

\begin{figure}[!t]
\centering
\includegraphics[width=0.85\textwidth]{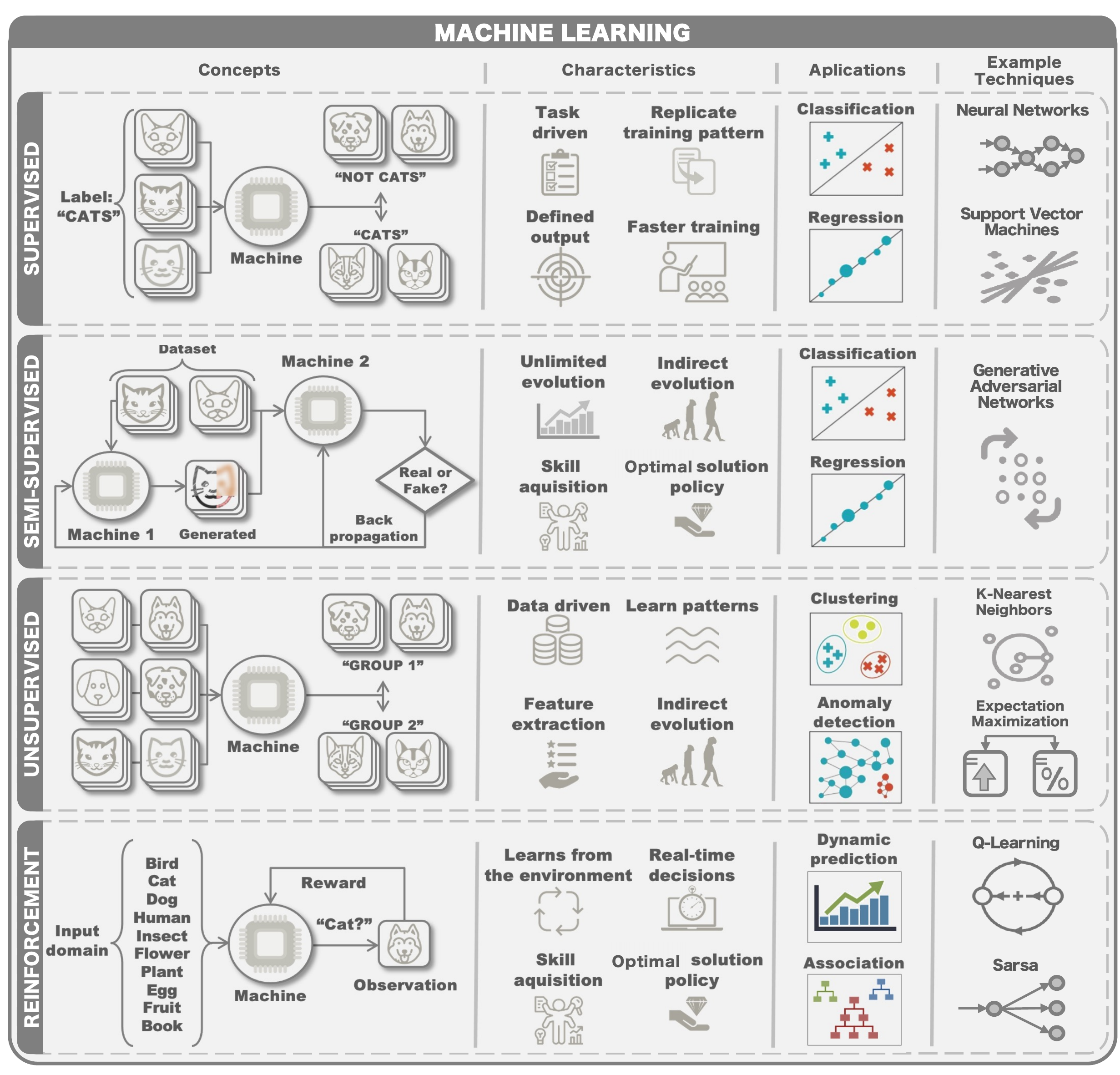}
\caption{Types of ML and their concepts, characteristics, and applications.}
\label{fig:MLsub}
\end{figure}

Machine learning (ML) constructs models from observations of data to make predictions~\cite{surveyMLinTesting2019}. Instead of being explicitly programmed like in traditional software, ML algorithms ``learn'' from observations using statistical analyses, facilitating the automation of decision making. 
ML has enabled many new applications in the past decade. As computational power and data availability increase, such approaches will increase in their capabilities and accuracy.

ML approaches largely fall into four categories---supervised, semi-supervised, unsupervised, and reinforcement learning---as presented in Figure~\ref{fig:MLsub}. In supervised learning, algorithms infer a model from the training data that makes predictions about newly encountered data. Such algorithms are typically used for classification---prediction of a label from a finite set---or regression---predictions in an unrestricted format, e.g., a continuous value. For example, a model may be trained from image data with the task of classifying whether an animal depicted in a new image is a cat. If a sufficiently large training dataset with a low level of noise is used, an accurate model can often be trained quickly. However, a model is generally static once trained and cannot be improved without re-training.

Semi-supervised algorithms are a form of supervised learning where feedback mechanisms are employed to automatically retrain models. For example, adversarial networks refine accuracy by augmenting the training data with new input by putting two supervised algorithms in competition. One of the algorithms creates new inputs that mimic training data, while the second predicts whether these are part of the training data or impostors. The first refines its ability to create convincing fakes, while the second tries to separate fakes from the originals. Semi-supervised approaches require a longer training time, but can achieve more optimal models, often with a smaller initial training set. 

Unsupervised algorithms do not use previously-labeled data. Instead, approaches identify patterns in data based on the similarities and differences between items. They model the data indirectly, with little-to-no human input. Rather than making predictions, unsupervised techniques aid in understanding data by, e.g., clustering related items, extracting interesting features, or detecting anomalies. As an example, a clustering algorithm could take a set of images and cluster them into groups based on the similarity of the images. Such an algorithm could not predict whether a specific image had a cat in it---as was done in supervised learning---but would likely place many of the cat-containing images in the same cluster. 

Reinforcement learning algorithms select actions based on an estimation of their effectiveness towards achieving a measurable goal~\cite{Almulla2020}. Reinforcement learning often does not require training data, instead learning through sequences of interactions with its environment. Reinforcement learning ``agents'' use feedback on the effect of actions taken to improve their estimation of the actions most likely to maximize achievement of their goal (their ``policy''). Feedback is provided by a reward function---a numeric scoring function. For example, an agent may predict which animal is contained in an image. It would then get a score based on how close its guess was to being correct---e.g., if the image contained a cat, then a guess of ``dog'' would get a higher score than a guess of ``spider''. The agent can also adapt to a changing environment, as estimations are refined each time an action is taken. Such algorithms are often the basis of automated processes, such as autonomous driving, and are effective in situations where sequences of predictions are required.

Recent research often focuses on ``deep learning''. 
Deep approaches make complex and highly accurate inferences from massive datasets. Many DL approaches are based on complex many-layered neural networks---networks that attempts to mimic how the human brain works~\cite{Li2019}. Such neural networks employ a cascade of nonlinear processing layers where one layer's output serves as the successive layer's input. 
Deep learning requires a computationally intense training process and larger datasets than traditional ML, but can learn highly accurate models, extract features and relationships from data automatically, and potentially apply models across applications. ``Deep'' approaches exist for all four of the ML types discussed above.

\subsection{Common Test Generation Techniques}\label{sec:test_gen_types}

Many techniques have been used to generate test input. In this subsection, we briefly introduce four common approaches: (a) random test generation, (b) search-based test generation, (c) symbolic execution, and (d), model-based test generation.

In random test generation, input is generated purely at random and applied to the system-under-test with the aim of triggering a failure. Random input generation is one of the most fundamental, simple, and easy-to-implement generation techniques~\cite{Anand13:Orchestrated}. Unfortunately, while random testing is often very efficient, most software has too large of an input space to exhaustively cover. Therefore, a weakness of random testing is that the generated input may only span a small, and uneven, portion of that input space. Therefore, many \textit{adaptive} random testing techniques have been proposed. In adaptive random testing, mechanisms are employed to partition the input space, and input is generated for each partition~\cite{Walkinshaw2017}. This ensures an even distribution across the input space. 

Search-based test generation formulates input generation as an optimization problem~\cite{Almulla2020}.  Of that near-infinite set of inputs for an SUT, we want to identify---systematically and at a reasonable cost---those that meet our goals. Given scoring functions that measure closeness to the attainment of those goals---fitness functions---metaheuristic optimization algorithms can automate that search by selecting input and measuring their fitness. Metaheuristic algorithms are often inspired by natural phenomena. For example, genetic algorithms evolve a population of candidate solutions over many generations by promoting, mutating, and breeding fit solutions. Such techniques retain many of the benefits of random testing, including scalability, and are often better able to identify failure-inducing input~\cite{Gay19:fitness}, or input with other properties of interest~\cite{Rojas15:Combining}.  

Symbolic execution is a program analysis technique where symbolic input is used instead of concrete input to ``execute'' the program~\cite{Luo2021}. The values of program variables are represented by symbolic expressions over these inputs. Then, at any point during a symbolic execution, the program's state can be represented by these symbolic values of program variables and a Boolean formula containing the collected constraints that must be satisfied for that path through the program to have been taken, also known as the path constraint. By identifying concrete input that satisfies a path constraint, we can ensure that particular paths through the program are covered by test cases. Constraint solvers can be used to identify such input automatically. Recent approaches often are based on dynamic symbolic execution (or concolic execution), where the symbolic execution is combined with concrete random execution to ease the difficulty of solving complex path constraints~\cite{Anand13:Orchestrated}. 

Finally, in model-based testing, lightweight models representing aspects of interest of an SUT are used to derive test cases~\cite{Shrestha2020a}. Often, such models take the form of a state machine. The internal behavior of the SUT is represented by its state. Transitions are triggered by applying input to the SUT. The model describes---at a chosen level of abstraction---the expected SUT behavior over a sequence of input actions. Test cases can then be derived from this model by choosing relevant subsets of input sequences~\cite{Anand13:Orchestrated}. 

\subsection{Related Work}\label{sec:related}

Other secondary studies overlap with ours in scope. We briefly discuss these publications below. Our SLR is the first focused specifically on the application of ML to automated test generation, including both input and oracle generation, and no related study overlaps in full with our research questions. We have also examined a larger and more recent sample of publications. 

Durelli et al. performed a systematic mapping study on the application of ML to software testing~\cite{surveyMLinTesting2019}. Their scope is broad, examining how ML has been applied to any aspect of the testing process. They mapped 48 publications to testing activities, study types, and ML algorithms employed. They observe that ML has been used to generate input and oracles. They note that supervised algorithms are used more often than other ML types and that Artificial Neural Networks are the most used algorithm. Jha and Popli also conducted a short review of literature applying ML to testing activities~\cite{Jha21:Review}, and note that ML has been used for both input and oracle generation. 

Ioannides and Eder conducted a survey on the use of AI techniques to generate test cases targeting code coverage---known as ``white box'' test generation~\cite{Ioannides12:CoverageReview}. Their survey focuses on optimization techniques, such as genetic algorithms, but they note that ML has been used to generate test input. 

Barr et al. performed a survey on test oracles~\cite{testOracleSurvey2014}. They divide test oracles into four types, including those specified by humans, those derived automatically, those that reflect implicit properties of programs, and those that rely on a human-in-the-loop. Approaches based on ML fall into the ``derived'' category, as they learn automatically from project artifacts to replace or augment human-written oracles. They discuss early approaches to using ML to derive oracles.

Balera et al. conducted a systematic mapping study on hyper-heuristics in search-based test generation~\cite{SBSTsurvey2019}. Search-based test generation applies optimization algorithms to generate test input. A hyper-heuristic is a secondary optimization performed to tune the primary search strategy, e.g., a hyper-heuristic could adapt test generation to the current SUT. A hyper-heuristic can apply ML, especially RL, but can also be guided by other algorithms. We also observe the use of ML-based hyper-heuristics. 

\section{Methodology}\label{sec:methodology}

Our aim is to understand how researchers have integrated ML into automated test generation, including generation of input and oracles. We have investigated publications related to this topic and seek to understand their methodology, results, and insights. To gain this understanding, we performed a systematic mapping study according to the guidelines of Petersen et al.~\cite{petersen2015guidelines}. 

We are interested in assessing the \textit{effect} of integrating ML into the test generation process, understanding the \textit{adoption} of these techniques---how and why they are being integrated, and which specific techniques are being applied, and identifying the potential \textit{impact} and \textit{risks} of this integration. Table~\ref{tab:RQ} lists the research questions we are interested in answering and clarifies the purpose of asking such questions. 

\begin{table}[!t]
\centering
\begin{tabular}{lll} 
\hline
\textbf{\textit{ID}}    & \textbf{Research Question}                                                                                                                  & \textbf{Objective}   \\ 
\hline
\textit{\textbf{ RQ1} } & \begin{tabular}[c]{@{}l@{}}Which testing practices have been supported by\\ integrating ML into the generation process? \end{tabular}        & \begin{tabular}[c]{@{}l@{}}Highlights testing scenarios and systems types \\ targeted for ML-enhanced test generation. \end{tabular}                                              \\ 
\hline
\textit{\textbf{ RQ2} } & \begin{tabular}[c]{@{}l@{}}What is the goal of using machine learning as \\ part of automated test generation? \end{tabular}                 & \begin{tabular}[c]{@{}l@{}}To understand the reasons for applying ML\\  techniques to perform or enhance test generation. \end{tabular}                                                               \\ 
\hline
\textit{\textbf{ RQ3} } & \begin{tabular}[c]{@{}l@{}}What types of ML have been used to perform \\ or enhance automated test generation? \end{tabular}               & \begin{tabular}[c]{@{}l@{}}Identifies the type of ML applied, how it was \\ integrated into the generation process, and \\ how it was trained and validated. \end{tabular}                                                    \\ 
\hline
\textit{\textbf{ RQ4} } & \begin{tabular}[c]{@{}l@{}} Which specific ML techniques were used to\\ perform or enhance automated test generation? \end{tabular}     & \begin{tabular}[c]{@{}l@{}}Identify specific ML techniques used in the \\process, including type, learning method, and \\ selection mechanisms. \end{tabular}                         \\ 
\hline
\textit{\textbf{ RQ5} } & How is the test generation process evaluated?                                                                                                & \begin{tabular}[c]{@{}l@{}}Describe the evaluation of the ML-enhanced \\ test generation process, highlighting common \\ metrics and artifacts  (programs or datasets) used. \end{tabular}   \\ 
\hline
\textit{\textbf{ RQ6} } & \begin{tabular}[c]{@{}l@{}}What are the limitations and open challenges \\ in integrating ML into test generation? \end{tabular}  & \begin{tabular}[c]{@{}l@{}}Highlights the limitations of enhancing test \\ generation with ML and future research directions. \end{tabular}             \\ 
\hline
\end{tabular}
\caption{List of research questions, along with motivation for answering the question.}
\label{tab:RQ}
\end{table}

The first three questions are high-level questions that clarify how ML has enabled or enhanced test generation, why ML was applied, and which specific testing scenarios were targeted by the enhanced generation techniques. \textbf{RQ1} enables us to categorize publications in terms of specific testing practices. By ``testing practices'', we refer either to the code or interface level that testing is aimed at (e.g., unit or GUI testing) or to specialized forms of testing (e.g., performance testing).  To answer this question, we divide the sampled publications into categories based on the specific goals and targets of test generation. We did not start with pre-decided categories but analyzed and thematically grouped publications. \textbf{RQ2} is motivational, covering the authors' primary objectives. We are interesting in how the authors intended to use ML---e.g., to directly generate input, to enhance an existing test generation technique, or to identify weaknesses in a testing strategy. 

In contrast, \textbf{RQ3-5} are technical questions. \textbf{RQ3} examines the broad category of ML technique (i.e., supervised, unsupervised, semi-supervised, or reinforcement learning), as well as its training and validation processes. \textbf{RQ4} examined which specific ML techniques (e.g., backpropagation neural networks) were used to perform the generation task. \textbf{RQ5} focuses on how the test generation approach is evaluated, including metrics and types of systems tested. This can include both the generation framework as a whole, or the specific ML aspect of the framework. Finally, the last research question covers the limitations of the proposed approaches and open research challenges (\textbf{RQ6}).

To answer these questions, we have performed the following tasks:
\begin{enumerate}
    \item Formed a list of publications by querying publication databases (Section~\ref{sec:selection}).
    \item Filtered this list for relevance (Section~\ref{sec:filtering}).
    \item Extracted and classified data from each study, guided by properties of interest (Section~\ref{sec:collection}).
    \item Identified trends in the extracted data to answer each research question (described along with results in Section~\ref{sec:results}).
\end{enumerate}

\subsection{Initial Study Selection}\label{sec:selection}

To locate publications for consideration, a search was conducted using four databases: IEEE Xplore, ACM Digital Library, Science Direct, and Scopus. To narrow the results, we created a search string by combining terms of interest on test generation and machine learning. The search string used was:
\begin{center}
\textit{(``test case generation'' OR ``test generation'' OR ``test oracle'' OR ``test input'') AND (``machine learning'' OR ``reinforcement learning'' OR ``deep learning'' OR ``neural network'')}
\end{center}

These keywords are not guaranteed to capture all publications on ML in test generation. However, they are intended to attain a relevant sample. Specifically, we combine terms related to test generation and terms related to machine learning, including common technologies. Our focus is not on any particular form of test generation. To obtain a representative sample, we have selected ML terms that we expect will capture a wide range of publications. These terms may omit some in-scope ML techniques, but attain a relevant sample while constraining the amount of manual inspection.

\begin{figure}[!t]
\centering
\includegraphics[width=0.75\textwidth]{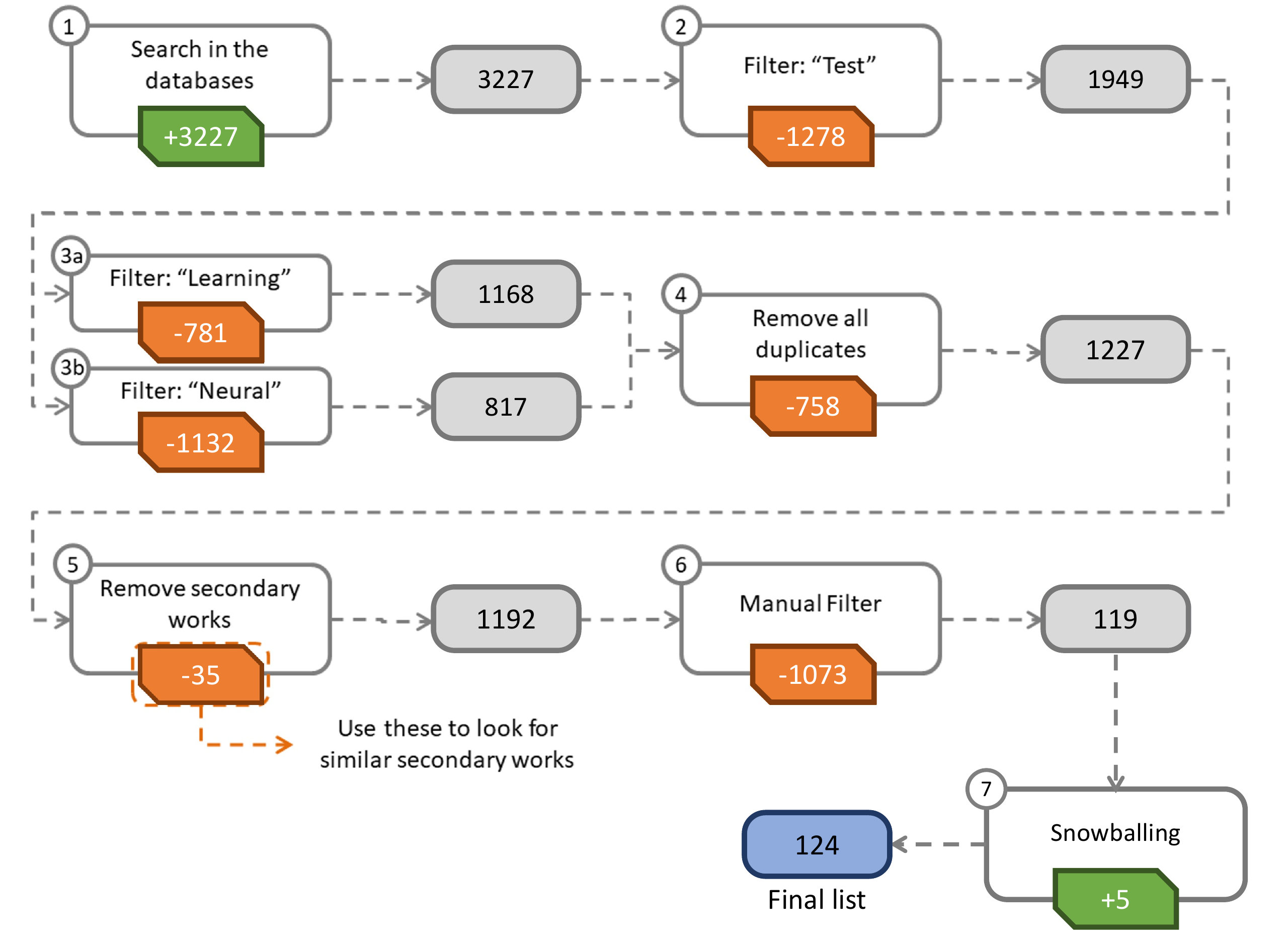}
\caption{Steps taken to determine the final list of publications to analyze.}
\label{fig:filtersEntries}
\end{figure}

We limited our search to peer-reviewed publications in English. The search string was applied to the full text of articles. Our set of articles was gathered in March 2023, containing an initial total of 3227 articles. This is shown as the first step in Figure~\ref{fig:filtersEntries}. 

To evaluate the search string's effectiveness, we conducted a verification process. First, we randomly sampled ten entries from the final publication list. Then we looked in each article for ten citations that were in scope, resulting in 100 citations. We checked whether the search string also retrieved these citations, and all 100 were retrieved. Although this is a small sample, it indicates the robustness of the string. 


\subsection{Selection Filtering}\label{sec:filtering}

We next applied a series of filtering steps to obtain a focused sample. Figure~\ref{fig:filtersEntries} presents the filtering process and the number of entries after applying each filter. The number in box 1 represents the initial number of articles. The numbers in the other boxes represent the number of entries removed in that particular step. The numbers between the steps show the total number of articles after applying the previous step. 

To ensure that publications are relevant, we used keywords to filter the list. We first searched the title and abstract of each study for the keyword ``test'' (including, e.g., ``testing''). We then searched the title and abstract of the remaining publications for either ``learning'' or ``neural''---representing application of ML. We merged the filtered lists, and removed all duplicate entries. We then removed all secondary studies. This left 1192 publications. 

\begin{figure}[!t]
\centering
\includegraphics[width=0.8\textwidth]{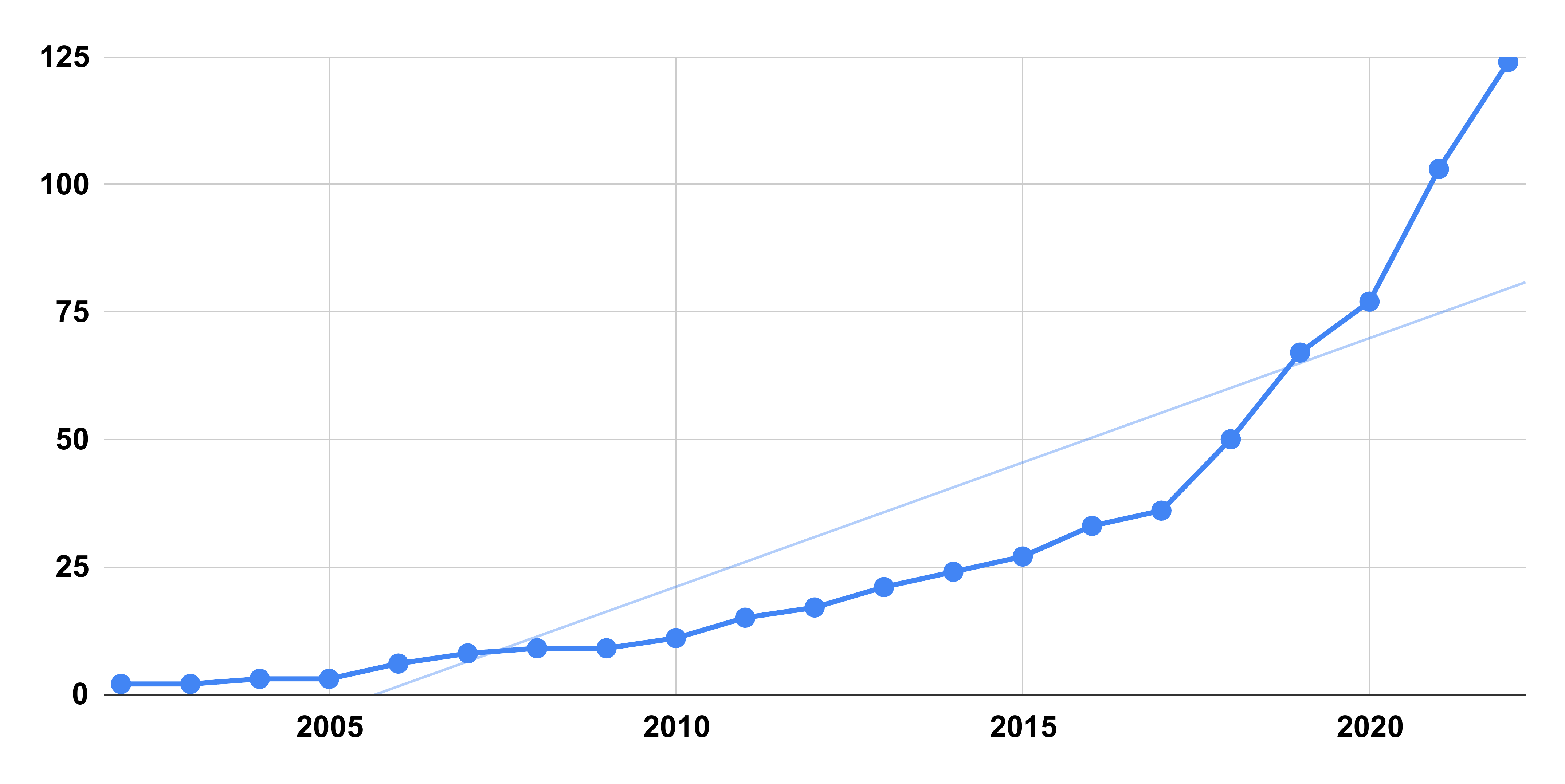} \vspace{-10pt}
\caption{Growth of the use of ML in test generation since 2002.}
\label{fig:articlesPerYearOverall}
\end{figure}

We examined the remaining publications manually, removing all publications not in scope. Publications were \textbf{included} if they met the following conditions:
\begin{itemize}
    \item The publication must be written in English.
    \item The publication must have appeared in a peer-reviewed venue. This includes journals, conferences, and workshops.
    \item The publication is a primary study (i.e., reporting original research). 
    \item The research reported relates to test generation---including test inputs, oracles, or full test cases---\textbf{and} applies any machine learning technique as part of the generation process. 
\end{itemize}
\noindent Articles were \textbf{excluded} under the following conditions:
\begin{itemize}
    \item The publication was not written in English.
    \item The publication was not in a peer-reviewed venue (e.g., book chapters, letters, white or grey literature).
    \item The publication is a secondary study (e.g., systematic literature review or mapping study). However, such studies were considered for discussion in Section~\ref{sec:related}. 
    \item The reported research does not relate to test generation, or the research is not discussed in the context of test generation\footnote{For example, ML could be applied as part of test suite reduction. Test suite reduction can be applied as part of a broader test generation framework, or it can be applied as a standalone testing technique. If the research was presented explicitly as part of test generation, we retained the publication. If the research was presented in a standalone context, then it was discarded.}. 
    \item The reported research does not apply ML as part of test generation (i.e., a non-ML technique is applied or ML is applied to an activity unrelated to test generation). 
    \item The reported research relates to testing \textit{of} ML-based systems rather than test generation. 
\end{itemize}

\begin{table}[!t]
\centering
\begin{tabular}{llll} 
\hline
\textit{\textbf{ID}}   & \textbf{Property Name} & \textbf{RQ} & \textbf{Description}  \\  \hline
\textit{\textbf{P1}}   & Testing Practices Addressed  & RQ1, RQ2  & \begin{tabular}[c]{@{}l@{}}The specific type of testing scenarios or application domain focused \\ on by the approach. It helps to categorize the publications, enabling \\ comparison between contributions. \end{tabular} \\ \hline
\textit{\textbf{P2}}   & Proposed Research  & RQ2   & A short description of the approach proposed or research performed.   \\ \hline
\textit{\textbf{P3}}   & Hypotheses and Results  & RQ1, RQ3    & \begin{tabular}[c]{@{}l@{}}Highlights the differences between expectations and conclusions of \\  the proposed approach. \end{tabular}  \\ \hline
\textit{\textbf{P4}}   & ML Integration & RQ3   & \begin{tabular}[c]{@{}l@{}}Covers how ML techniques have been integrated into the test\\ generation process. It is essential to understand what aspects of \\ generation are handled or supported by ML. \end{tabular}   \\ \hline
\textit{\textbf{P5}}   & ML Technique Applied  & RQ4  & \begin{tabular}[c]{@{}l@{}}Name, type, and description of the ML technique used in the study. \end{tabular}  \\ \hline
\textit{\textbf{P6}}   & \begin{tabular}[c]{@{}l@{}}Reasons for Using the \\ Specific ML Technique \end{tabular} & RQ4   & \begin{tabular}[c]{@{}l@{}}The reasons stated by the authors for choosing this ML technique.  \end{tabular}  \\ \hline
\textit{\textbf{P7}}   & ML Training Process  & RQ4  & \begin{tabular}[c]{@{}l@{}}How the approach was trained, including the specific data sets or \\ artifacts used to perform this training. This property helps us \\ understand how each contribution could be replicated or extended. \end{tabular}   \\ 
\hline
\textit{\textbf{P8}}   & \begin{tabular}[c]{@{}l@{}}External Tools or \\ Libraries Used \end{tabular}  & RQ4 & \begin{tabular}[c]{@{}l@{}} External tools or libraries used to implement the ML technique. \end{tabular}  \\ \hline
\textit{\textbf{P9}}   & \begin{tabular}[c]{@{}l@{}} ML Objective and \\Validation Process  \end{tabular}& RQ4, RQ5    & \begin{tabular}[c]{@{}l@{}}This attribute covers the objective of the ML technique (e.g., \\ reward function or validation metric), and how it is \\ validated, including data, artifacts, and metrics used (if any). \end{tabular}   \\ \hline
\textit{\textbf{P10}}  & \begin{tabular}[c]{@{}l@{}}Test Generation \\ Evaluation Process \end{tabular} & RQ5         & \begin{tabular}[c]{@{}l@{}}
Covers how the ML-enhanced oracle generation process, as a \\
whole, is evaluated (i.e., how successful are the generated \\ 
input at triggering faults or meeting some other testing goal?).\\ 
Allows understanding of the effects of ML on improving the \\ testing process.
\end{tabular}  \\ \hline
\textit{\textbf{P11}}  & Potential Research Threats  & RQ6 & Notes on the threats to validity that could impact each study.  \\ \hline
\textit{\textbf{P12}}  & Strengths and Limitations  & RQ6 & \begin{tabular}[c]{@{}l@{}}This property is used to understand the general strengths and \\ limitations of enhancing a generation process with ML by \\ collecting and synthesizing these aspects for both the ML \\ techniques  and entire test generation approaches. \end{tabular} \\ \hline
\textit{\textbf{P13}}  & Future Work & RQ6 & \begin{tabular}[c]{@{}l@{}}Any future extensions proposed by the authors, with a particular \\ focus on those that could overcome the identified limitations. \end{tabular}  \\ \hline
\end{tabular}
\caption{List of properties used to answer the research questions. For each property, we include a name, the research questions the property is associated with, and a short description.}
\label{tab:props}
\end{table}

This determination was made by first reading the title, abstract, and introduction. Then, if the publication seemed in scope, we proceeded to read the entire study. In a small number of cases, publications were deemed out-of-scope only after inspection of the full article. 
Both authors independently inspected publications during this step to prevent the accidental removal of relevant publications. In cases of disagreement, the authors discussed the study. 



This process resulted in a sample of \papersinitial articles. We then performed snowballing by inspecting the bibliography of each publication and adding any additional publications that met our inclusion criteria stated above. The snowballing process added \paperssnowballing additional publications, resulting in a final sample of \papers publications. 

The publications are listed in Section~\ref{sec:by-area}, associated with the specific testing practice addressed. Figure~\ref{fig:articlesPerYearOverall} shows the growth of interest in this topic since 2002 (only one study in this sample, from 1993, was published before this date). We can see modest, but growing, interest until 2010. The advancements in ML in the past decade have resulted in significantly more use of ML in test generation, especially starting in 2018. Over 70\% of the publications in our sample were published in the past five years alone---with 38\% in the past two years. This is an area of growing interest and maturity, and we expect the number of publications to increase significantly in the next few years.

\subsection{Data Extraction and Classification}\label{sec:collection}

To answer the questions in Table~\ref{tab:RQ}, we have extracted a set of key properties from each study, identified in Table~\ref{tab:props}. Each property listed in the table is briefly defined and is associated with the research questions. Several properties may collectively answer a RQ. For example, RQ2---covering the goals of using ML---can be answered using property P2. However, P1 provides context and the testing practice addressed may dictate how ML is applied.

Data extraction was performed primarily by the first author of this study. However, to ensure the accuracy of the extraction process, the second author performed a full independent extraction for a sample of ten randomly-chosen publications. We compared our findings, and found that we had near-total agreement on all properties. The second author then performed a lightweight verification of the findings of the first author for the remaining publications. A small number of corrections were discussed between the authors, but the data extraction was generally found to be accurate. 

Systematic mapping studies generally address research questions by grouping publications into different \textit{classifications}, then analyzing trends in the publications in each group. We likewise group publications in the following ways: 
\begin{itemize}
    \item The testing practice addressed. We first divide the research into input and oracle generation, then a specific input granularity (e.g., unit or system-level input generation) or input/test type (e.g., performance testing) or specific oracle type (e.g., test verdicts, expected output). 
    \item The type of ML applied (e.g., supervised or reinforcement learning). 
    \item The specific ML technique applied (e.g., backpropagation neural network).
    \item The type of training data used, if applicable (e.g., previous system executions). 
    \item The objective of applying ML. This includes both the type of prediction being made (e.g., classification or regression) and the purpose of the prediction (e.g., predicting the input that will cover a path in the code). For reinforcement learning, this includes the reward functions used. 
    \item The evaluation metrics used to assess the proposed research. This includes both traditional test generation evaluation metrics (e.g., number of faults detected) and ML-related metrics (e.g., accuracy). 
    \item The type of example systems used in the evaluation. 
\end{itemize}

For the ML approach, we use the four primary categories of ML described in Section~\ref{sec:background} to classify publications---supervised, semi-supervised, unsupervised, and reinforcement learning. For the other categories, we did not begin with pre-determined classifications. Rather, we performed thematic analysis of the articles to identify natural groupings in the publication sample. In all cases, our goal was to avoid oversimplification---we favored a large number of specialized classes over a small number of over-arching classes. We describe classification more concretely in Section~\ref{sec:results}.

\section{Results and Discussion}\label{sec:results}

In this section, first, we identify the testing practices addressed by ML-enhanced test generation (RQ1, Section~\ref{sec:problems}). We then note observations regarding research related to individual testing practices (Section~\ref{sec:by-area}). Finally, we present answers to RQ2-6 (Sections~\ref{sec:rq2}--\ref{sec:rq6}). 

\subsection{RQ1: Testing Practices Addressed}
\label{sec:problems}

\begin{figure}[!t]
\centering
\includegraphics[width=.76\textwidth]{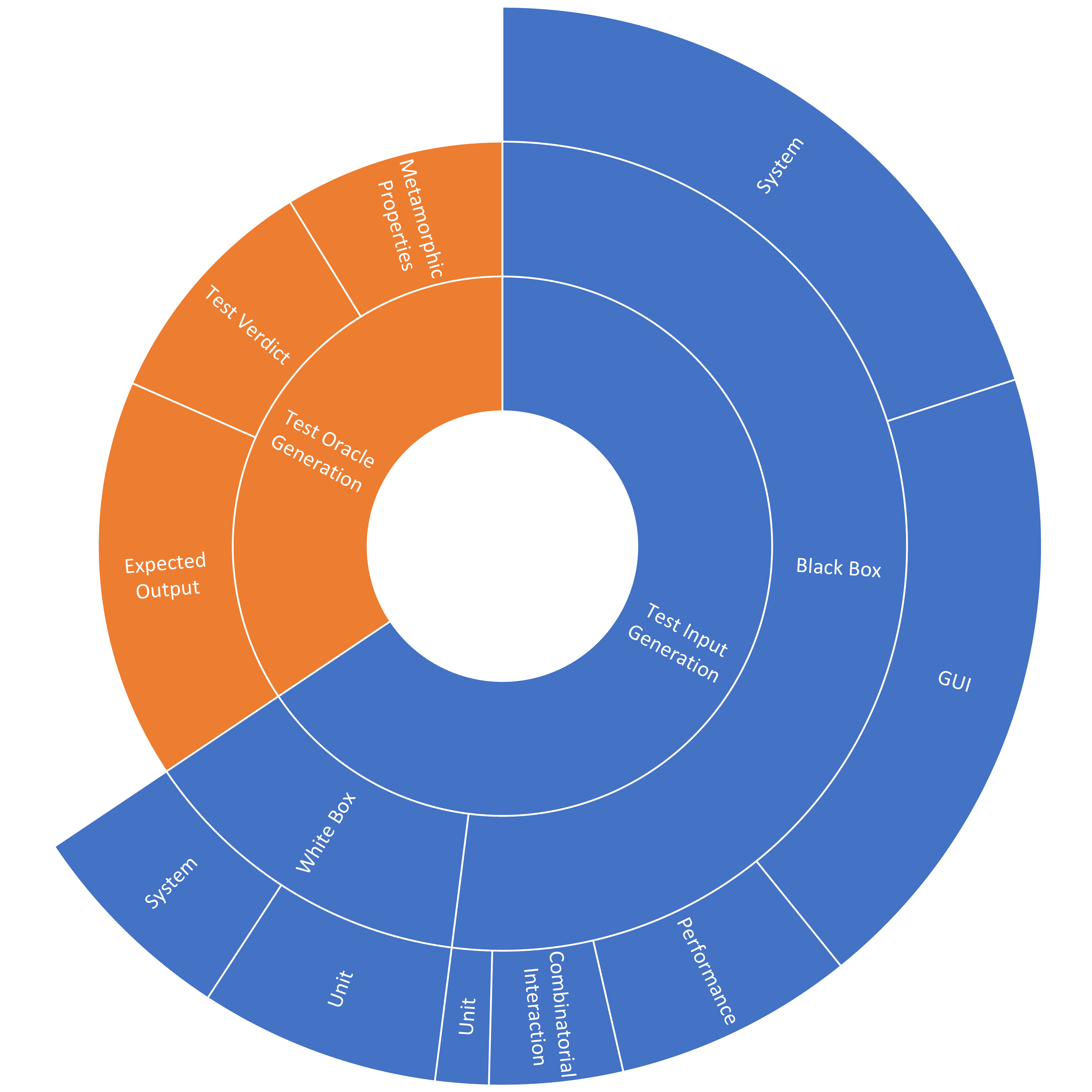}
\resizebox{\columnwidth}{!}{
\begin{tabular}{lcllcllc}
\multicolumn{2}{c}{\textbf{Layer \#1}} & \multicolumn{1}{c}{\textbf{}} & \multicolumn{2}{c}{\textbf{Layer \#2}} & \multicolumn{1}{c}{\textbf{}} & \multicolumn{2}{c}{\textbf{Layer \#3}}  \\ \cline{1-2} \cline{4-5} \cline{7-8} 
Test Input Generation        & \papersinput      &                               & Black Box                    & \papersblackbox      &                               & System Test Generation (Black Box)             & \paperssystemblack \\
Test Oracle Generation       & \papersoracle      &                               & White Box                    & \paperswhitebox      &                               & GUI Test Generation                & \papersgui \\ \cline{1-2}
                             &         &                               & Expected Output              & \papersexpected      &                               & Performance Test Generation        & \papersperformance  \\
                             &         &                               & Test Verdict                 & \papersverdict      &                               & Combinatorial Interaction Testing  & \paperscit \\
                             &         &                               & Metamorphic Properties      & \papersmetamorphic      &                               & Unit Test Generation (Black Box)   & \papersunitblack  \\ \cline{4-5}
                             &         &                               &                              &         &                               & Unit Test Generation (White Box)   & \papersunitwhite  \\
                             &         &                               &                              &         &                               & System Test Generation (White Box) & \paperssystemwhite  \\ \cline{7-8} 
\end{tabular}
}
\caption{Testing practices addressed by test generation approaches incorporating ML.}
\label{fig:articleDivision} \vspace{10pt}
\end{figure}

The purpose of RQ1 is to give an overview of which testing practices have been targeted by the publications to help structure our examination of the sampled articles. Our categorization is shown in Figure~\ref{fig:articleDivision}. In this chart, we divide articles into layers, with each layer representing finer levels of granularity. The total number of publications in each category is reported below. 

The specific formulation of a test case depends on the product domain and technologies utilized by the SUT~\cite{Pezze06:testing}. However, broadly, a test case is defined by a set of input steps and test oracles~\cite{testOracleSurvey2014}, both of which can be the target of automated generation. Therefore, we decided that \textit{input} and \textit{oracles} constitute our first division.

A majority of articles focus on input generation (67\% of the sample). Automated input generation has become a major research topic in software testing over the past 20 years~\cite{Orso14:STR}, and many different forms of automated generation have been proposed, using approaches ranging from symbolic execution~\cite{Luo2021} to optimization~\cite{Almulla2020}. Oracle generation has long been seen as a major challenge for test automation research~\cite{testOracleSurvey2014,Orso14:STR}. However, ML is a realistic route to achieve automated oracle generation~\cite{Fontes21:SLR}, and a significant number of publications have started to appear on this topic (33\%).

\begin{figure}[!t]
\centering
   \begin{subfigure}[t]{\textwidth}
        \centering
        \includegraphics[width=0.73\textwidth]{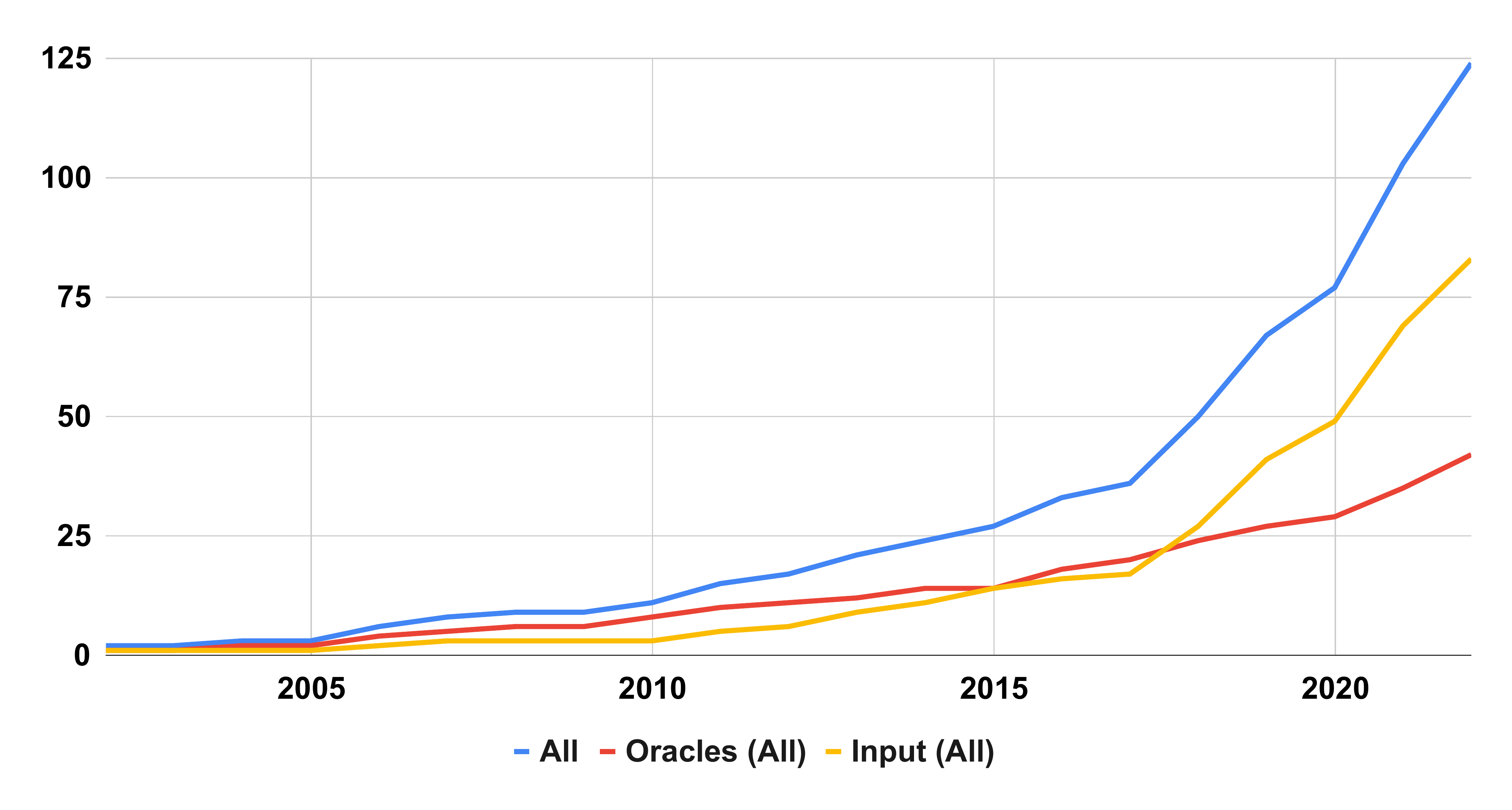} \vspace{-10pt}
        \caption{Input and Oracle Generation}
    \end{subfigure}
    
    \begin{subfigure}[t]{\textwidth}
        \centering
        \includegraphics[width=0.73\textwidth]{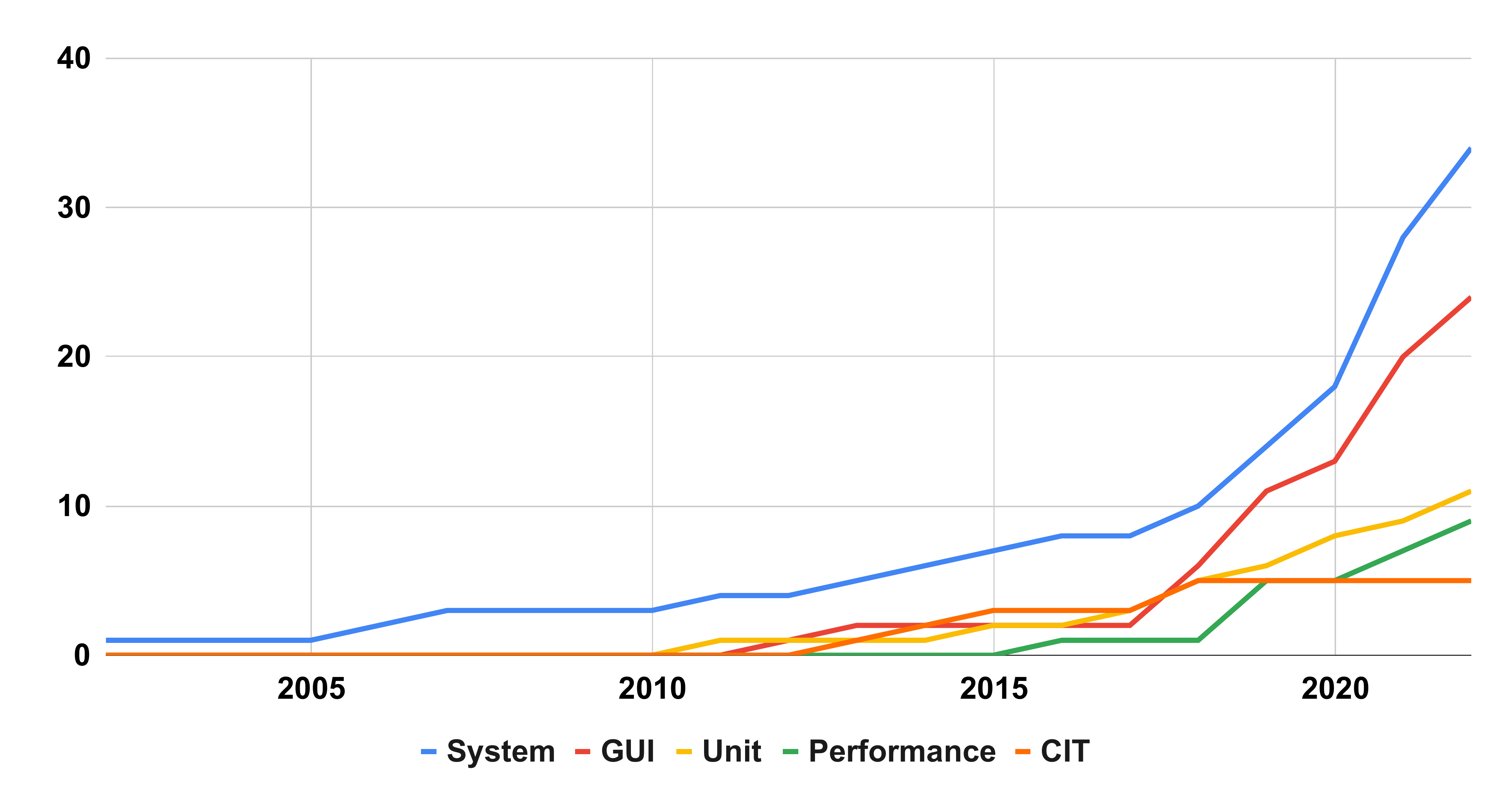} \vspace{-10pt}
        \caption{Forms of Input Generation}
    \end{subfigure}
    
    \begin{subfigure}[t]{\textwidth}
        \centering
        \includegraphics[width=0.73\textwidth]{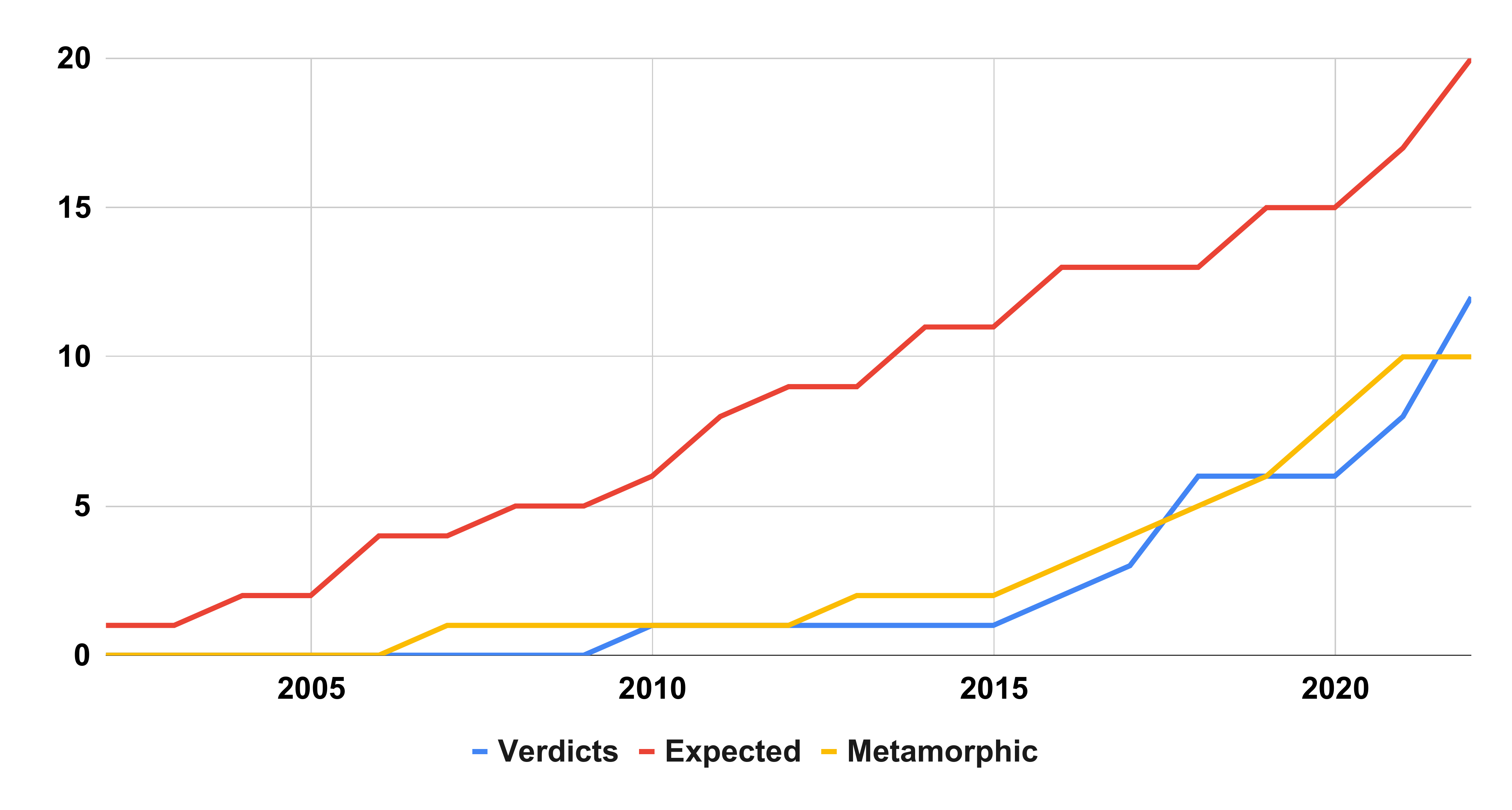}
        \caption{Forms of Oracle Generation}
    \end{subfigure}
\caption{Grown in the use of ML in test generation since 2002.}
  \label{fig:articlesPerYear}
\end{figure}

Figure~\ref{fig:articlesPerYear}(a) shows the growth in both topics since 2002. Both show a similar trajectory until 2017, with a sharp increase in input generation after. New ML technologies, such as deep learning, and the growing maturity of open-source learning frameworks, such as PyTorch, Keras, and OpenAI Gym, have potentially contributed to this increase. 


\subsubsection{Test Input Generation}

In the second layer of Figure~\ref{fig:articleDivision}, we further divided test input generation by the source of information used to create test input:
\begin{itemize}
    \item \textbf{Black Box Testing:} Also known as \textbf{functional testing}~\cite{Pezze06:testing}, approaches use information about the program gleaned from documentation, requirements, and other project artifacts to create test inputs.  
    \item \textbf{White Box Testing:} Also known as \textbf{structural testing}~\cite{Pezze06:testing}, approaches use the source code to select test inputs (e.g., generating input that covers a particular outcome for an \texttt{if}-statement). Approaches do not require domain knowledge. 
\end{itemize}
Of the \papersinput publications addressing input generation, \papersblackbox propose Black Box and \paperswhitebox propose White Box approaches. White Box approaches are traditionally common in input generation, as the ``coverage criteria''---checklists of goals~\cite{Majma2014}---that are the focus of White Box testing offer measurable test generation targets~\cite{Almulla2020}. Such approaches benefit from the inclusion of ML~\cite{Almulla2020}. However, ML may have great potential to enhance Black Box testing. Black Box approaches are based on external data about how the system should behave. ML can be used to automate analyses of that data---enabling new approaches to test generation---as shown by 80\% of input generation publications proposing Black Box approaches.

\noindent In the third layer of Figure~\ref{fig:articleDivision}, we further subdivided approaches based on either the level of granularity that generated inputs were applied at or by the specialized form of input generated:
\begin{itemize}
    \item \textbf{System Test Generation (\paperssystem publications):} A practice where tests target a subsystem or full system through a defined interface (e.g., API or CLI) and verify high-level functionality through that interface. 
    \item \textbf{GUI Test Generation (\papersgui publications):} A specialized form of system testing where tests target a GUI to identify incorrect functionality or usability/accessibility issues~\cite{Ariyurek2021}. We also incorporate game testing (when conducted through a GUI) into this category~\cite{Zheng19:Game}. 
    \item \textbf{Unit Test Generation (\papersunit publications):} A practice where test cases target a single class and exercise its functionality in isolation from other classes. 
    \item \textbf{Performance Test Generation (\papersperformance publications):} Tests are generated to assess whether the SUT meets non-functional requirements (e.g., speed, scalability, or resource usage requirements)~\cite{Ahmad2019}
    \item \textbf{Combinatorial Interaction Testing (\paperscit publications):} A system-level practice that attempts to produce a small set of tests that cover important interactions between input variables~\cite{Jia2015}. 
\end{itemize}

System-level testing is the most common category (41\% of input generation), followed by GUI (29\%), then unit testing (13\%). GUI, performance (11\%) and combinatorial interaction testing (CIT) (6\%) represent specialized forms of system testing. 

Figure~\ref{fig:articlesPerYear}(b) shows the growth in each area of input generation. We see a particularly strong growth in system and GUI testing since 2017. In addition to the emergence of open-source ML frameworks, we also hypothesize that this is partially driven by the emergence of mobile and web applications and autonomous vehicles. Mobile applications are tested primarily through a GUI, as are many web applications---leading to increased interest in GUI testing. Only two GUI test generation articles predate 2017, and of the post-2017 articles, 86\% relate to testing of either mobile or web applications (see Table~\ref{tab:GUIresults}). Other web applications are tested through REST APIs, and are included in system testing. Autonomous vehicles also require new approaches, as they are tested in complex simulators~\cite{Baumann2021}. Since 2017, web and autonomous vehicle testing constitute the two largest dedicated domains for system testing, with 15\% and 19\% of post-2017 publications, respectively (see Tables~\ref{tab:systemTesting}-\ref{tab:systemTestingWhite}). 


\subsubsection{Test Oracle Generation}


The second layer under test oracle generation in Figure~\ref{fig:articleDivision} divides approaches based on the \textit{type} of test oracle produced:
\begin{itemize}
    \item \textbf{Expected Output (\papersexpected publications):} The oracle predicts concrete output behavior that should result from an input. Often, this will be abstracted (i.e., a \textit{class} of output). 
    \item \textbf{Test Verdicts (\papersverdict publications):} The oracle predicts the final test verdict for a given input (i.e., a ``pass'' or ``fail'').
    \item \textbf{Metamorphic Relations and Other Properties of Program Behavior (\papersmetamorphic publications):} A metamorphic relation is a property relating input to expected output~\cite{Hardin2018}---e.g., $sin(x) = sin(\pi - x)$. Such properties, as well as other property types that specify the expected behavior of a SUT, can be applied to many inputs. Violations of such properties identify potential faults. 
\end{itemize}

ML supports decision processes. A ML technique makes a prediction, which can either be a decision or information that supports making a decision. Test oracles follow a similar model, consisting of \textit{information} used to issue a verdict and a \textit{procedure} to arrive at a verdict~\cite{Gay15:oracleselection}. ML offers a natural means to replace either component. Test verdict oracles replace the procedure, while expected output and property oracles support arriving at a verdict. Figure~\ref{fig:articlesPerYear}(c) shows steady growth for all types. 

\begin{center}
\begin{framed}
  \textbf{RQ1 (Testing Practices):} ML supports the generation of both test input and oracles, with a greater focus on input generation (67\% of the sample). Input generation research targets system testing, specialized types of system testing (GUI, performance, CIT), and unit testing. The majority of these are Black Box approaches, with White Box approaches primarily restricted to unit testing. There has been an increase in system and GUI input generation since 2017, potentially related to the emergence of web and mobile applications and autonomous driving, as well as to the availability of robust, open-source ML and deep learning frameworks. ML supports generation of test verdict, metamorphic (and other property-based), and expected output oracles.
\end{framed}
\end{center}


\subsection{Examining Specific Practices}\label{sec:by-area}

Before answering the remaining research questions, we examine concretely how ML has supported test generation. 

\begin{table}[!t]
\resizebox{\columnwidth}{!}{
\begin{tabular}{llllllll}
\hline
\textbf{Ref} & \textbf{Year} & \textbf{ML Approach} & \textbf{Technique} & \textbf{Training Data}  & \textbf{ML Objective} & \textbf{Evaluation Metrics} & \textbf{Evaluated On} \\ \hline
\cite{AraizaIllan2016}  & 2016 & Reinforcement& Q-Learning  & N/A & \begin{tabular}[c]{@{}l@{}}Reward (Plan Coverage)\end{tabular} & \begin{tabular}[c]{@{}l@{}}Code Coverage, \\ Assertion Coverage\end{tabular} & Robotic Systems \\ \hline
\cite{Baumann2021} & 2021 & Reinforcement& Q-Learning & N/A  & \begin{tabular}[c]{@{}l@{}}Reward (Criticality)\end{tabular} & Faults Detected  & Autonomous Vehicles \\ \hline
\cite{Buzdalov2013a}    & 2013 & Reinforcement & \begin{tabular}[c]{@{}l@{}}Delayed Q-Learning\end{tabular} & N/A  & \begin{tabular}[c]{@{}l@{}}Reward (Test Improvement)\end{tabular} & \begin{tabular}[c]{@{}l@{}}\% of Runs Where \\ Requirements Met\end{tabular} & Ship Logistics \\  \hline
\cite{Esnaashari2021}  & 2021 & Reinforcement& Q-Learning & N/A  & \begin{tabular}[c]{@{}l@{}}Reward (Code Coverage)\end{tabular}  & Code Coverage & \begin{tabular}[c]{@{}l@{}}Triangle Classification, \\ Nesting Structure,\\ Complex Conditions\end{tabular} \\ \hline
\cite{Huurman2020}   & 2020 & Reinforcement & Asynchronous Advantage Actor Critic & N/A  & \begin{tabular}[c]{@{}l@{}}Reward (Transition Coverage)\end{tabular}  & Not Evaluated & OpenAPI APIs \\ \hline
\cite{Reddy2020}   & 2020  & Reinforcement& \begin{tabular}[c]{@{}l@{}}Monte Carlo Control\end{tabular} & N/A & \begin{tabular}[c]{@{}l@{}}Reward (Input Diversity)\end{tabular}  & \begin{tabular}[c]{@{}l@{}}Input Diversity, \\ Code Coverage\end{tabular} & \begin{tabular}[c]{@{}l@{}}XML, JavaScript \\ Parsing\end{tabular} \\ \hline
\cite{Shu22:EFSM} & 2022 & Reinforcement & Deep Q-Network & N/A & Reward (Transition Coverage) & \begin{tabular}[c]{@{}l@{}}Efficiency, \\ Sensitivity,\\ Transition Cov.\end{tabular} & \begin{tabular}[c]{@{}l@{}}State Machine\\Benchmark\end{tabular}\\ \hline 
\cite{Veanes06:RL} & 2006 & Reinforcement & Markov Decision Process & N/A & Reward (State Coverage) & State Coverage & Recycling Robot \\ \hline
\cite{Deng2021}  & 2021 & Semi-supervised & \begin{tabular}[c]{@{}l@{}}Generative Adversarial Network, \\Convolutional NN\end{tabular}  & Image Input  & \begin{tabular}[c]{@{}l@{}}Regression (Speed)\end{tabular}& Faults Detected & Autonomous Vehicles \\ \hline
\cite{Zhang18:DeepRoad} & 2018 & Semi-supervised & \begin{tabular}[c]{@{}l@{}}Generative Adversarial Network\end{tabular}  & Image Input  & \begin{tabular}[c]{@{}l@{}}Regression (Steering Angle)\end{tabular}& \begin{tabular}[c]{@{}l@{}}Input Validity, \\ Faults Detected\end{tabular} & Autonomous Vehicles \\ \hline
\cite{Bergadano1993a}   & 1993  & Supervised  & Not Specified  & System Executions  & \begin{tabular}[c]{@{}l@{}}Regression (Output)\end{tabular} & Not Evaluated & N/A \\ \hline
\cite{Budnik2018}  & 2018 & Supervised  & Backpropagation NN & System Executions & \begin{tabular}[c]{@{}l@{}}Regression (Output)\end{tabular}  & Output Coverage & Train Controller \\ \hline
\cite{Eidenbenz2021} & 2021 & Supervised & \begin{tabular}[c]{@{}l@{}}Gaussian Process, Decision Trees, \\ AdaBoostedTree, Random Forest, \\ Support Vector Machine, Artificial NN\end{tabular} & System Executions  & \begin{tabular}[c]{@{}l@{}}Regression (Output)\end{tabular} & Accuracy & Power Grid Control  \\ \hline
\cite{Gao2019}   & 2019   & Supervised & Long Short-Term Memory NN & Existing Inputs & \begin{tabular}[c]{@{}l@{}}Regression (Valid  Input)\end{tabular} & \begin{tabular}[c]{@{}l@{}}Accuracy, \\ Code Coverage\end{tabular}  & \begin{tabular}[c]{@{}l@{}}FTP Programs\end{tabular} \\ \hline
\cite{Kikuma2019} & 2019 & Supervised  & Conditional Random Fields & Test Descriptions & \begin{tabular}[c]{@{}l@{}}Regression (Requirement\\ Associations)\end{tabular}  & Accuracy & Telecom Systems \\ \hline
\cite{Kirac2019} & 2019 & Supervised  & Long Short-Term Memory NN & Existing Inputs & \begin{tabular}[c]{@{}l@{}}Regression (Failing Input)\end{tabular}  & \begin{tabular}[c]{@{}l@{}}Faults Detected, \\ Efficiency\end{tabular} & Smart TV \\ \hline 
\cite{Meinke2021}  & 2021 & Supervised & \begin{tabular}[c]{@{}l@{}}Parallel Distributed \\ Processing\end{tabular} & System Executions & \begin{tabular}[c]{@{}l@{}}Regression (Output)\end{tabular}  & \begin{tabular}[c]{@{}l@{}}Efficiency, \\ Faults Detected, \\ Model Size \end{tabular}  & Autonomous Vehicles  \\ \hline
\cite{Mirabella2021}  & 2021 & Supervised & Multilayer Perceptron & System Executions & \begin{tabular}[c]{@{}l@{}}Classification (Input Validity)\end{tabular}  & Accuracy & \begin{tabular}[c]{@{}l@{}}REST APIs (GitHub, \\ LanguageTool, Stripe, \\ Yelp, YouTube)\end{tabular}  \\ \hline
\cite{Sharma22:Property} & 2022 & Supervised & Regression Tree, Feedforward NN & System Executions & Regression (Output) & \begin{tabular}[c]{@{}l@{}}Efficiency, \\ Faults Detected\end{tabular}
& Numeric Functions \\ \hline
\cite{Shrestha2020a}  & 2020  & Supervised  & Long Short-Term Memory NN & Simulink models & \begin{tabular}[c]{@{}l@{}}Regression (Validity Rules)\end{tabular}  & \begin{tabular}[c]{@{}l@{}}Input Validity, \\ Faults Detected\end{tabular} & Simulink tools \\ \hline
\cite{Ueda2021} & 2021 & Supervised & Conditional Random Fields  &  Specifications  & \begin{tabular}[c]{@{}l@{}}Regression (Requirement \\ Associations)\end{tabular} & Accuracy & Unspecified  \\ \hline
\cite{Utting2020} & 2020 & \begin{tabular}[c]{@{}l@{}}Supervised, \\ Unsupervised\end{tabular}  & \begin{tabular}[c]{@{}l@{}}Decision Trees, Gradient Boosting, \\K-Nearest Neighbor, MeanShift\end{tabular} & System Executions & \begin{tabular}[c]{@{}l@{}}Regression (Validity Rules), \\Clustering (Covered Input)\end{tabular} & \begin{tabular}[c]{@{}l@{}}Num. Clusters, \\ Accuracy, \\ Event Coverage\end{tabular} & \begin{tabular}[c]{@{}l@{}}Bus System, \\ Supply Chain\end{tabular}  \\ \hline
\cite{Zhao2007} & 2007  & Supervised  & Backpropagation NN & System Executions & \begin{tabular}[c]{@{}l@{}}Regression (Output)\end{tabular} & \begin{tabular}[c]{@{}l@{}}Accuracy, Efficiency \end{tabular} & \begin{tabular}[c]{@{}l@{}}Fault Tolerant \\ System,  Arc \\ Length \end{tabular} \\ \hline
\cite{Zhong22:Fuzz} & 2022 & Supervised & Shallow NN & \begin{tabular}[c]{@{}l@{}}RNG Seeds, \\System Executions\end{tabular} & \begin{tabular}[c]{@{}l@{}}Regression (Prob. \\Traffic Violation)\end{tabular} & \begin{tabular}[c]{@{}l@{}}Adaptivity, \\ Faults Detected, \\ Sensitivity\end{tabular} & Autonomous Vehicles \\ \hline
\cite{Zhu2019}   & 2019  & Supervised  & Support Vector Machine  & Existing Inputs & \begin{tabular}[c]{@{}l@{}}Regression (Validity Rules)\end{tabular} & \begin{tabular}[c]{@{}l@{}}Tests Generated,\\ Tests Executed,\\ Test Size, \\ Faults Detected\end{tabular} & \begin{tabular}[c]{@{}l@{}}Domain-Specific \\ Compiler\end{tabular} \\ \hline
\end{tabular}}
\caption{
Publications under \textbf{System Test Generation (Black Box)} with publication date, ML type, ML technique, training data, objective of the ML, evaluation metrics, and applications used to evaluate. NN = Neural Network.
} 
\label{tab:systemTesting}
\end{table}

\subsubsection{System Test Generation}

\begin{table}[!t]
\resizebox{\columnwidth}{!}{
\begin{tabular}{llllllll}
\hline
\textbf{Ref} & \textbf{Year} & \textbf{ML Approach} & \textbf{Technique} & \textbf{Training Data}  & \textbf{ML Objective} & \textbf{Evaluation Metrics} & \textbf{Evaluated On} \\ \hline
\cite{Chen2021i} & 2021 & Reinforcement& ReLU Q-Learning & Constraints & \begin{tabular}[c]{@{}l@{}}Reward (Solving Cost)\end{tabular} & \begin{tabular}[c]{@{}l@{}}Code Coverage,\\ Queries Solved\end{tabular}  & GNU coreutils\\ \hline
\cite{Paduraru2021} & 2021 & Reinforcement& Deep Q-Network & N/A  & \begin{tabular}[c]{@{}l@{}}Reward (Code Coverage, \\Path Length)\end{tabular}& Code Coverage   & Sorting  \\ \hline
\cite{Feldmeier22:Games} & 2022 & Supervised &  Artificial NN & Game States & Regression (Input Action) & \begin{tabular}[c]{@{}l@{}}Code Coverage, \\ Mutation Score\end{tabular} & Scratch Games \\ \hline
\cite{Liu22:CovGen} & 2022 & Supervised & Radial-Basis Function NN & \begin{tabular}[c]{@{}l@{}}Existing Inputs, \\ Code Coverage\end{tabular} & Regression (Code Coverge) & Code Coverage & Numeric Functions \\ \hline
\cite{Luo2021} & 2021 & Supervised & \begin{tabular}[c]{@{}l@{}}Long Short-Term Memory NN, \\ Tree-LSTM, K-Nearest Neighbour\end{tabular}  & Constraints & \begin{tabular}[c]{@{}l@{}}Regression (Solving Time)\end{tabular}  & \begin{tabular}[c]{@{}l@{}}Accuracy, Constraint\\ Solving Time\end{tabular} & \begin{tabular}[c]{@{}l@{}}GNU coreutils, \\ Busybox utils,\\ SMT-COMP\end{tabular}\\ \hline
\cite{Majma2014} & 2014   & Supervised  & Backpropagation NN & \begin{tabular}[c]{@{}l@{}}Existing Inputs, \\ Code Coverage\end{tabular} & \begin{tabular}[c]{@{}l@{}}Regression (Code Coverage)\end{tabular}   & Code Coverage & \begin{tabular}[c]{@{}l@{}} Binary Search, \\ Sorting, \\ Median, GCD, \\ Triangle Class.\end{tabular} \\  \hline  
\cite{Mishra2011}  & 2011 & Supervised  & Backpropagation NN  &  \begin{tabular}[c]{@{}l@{}}Existing Inputs, \\ Code Coverage\end{tabular} & \begin{tabular}[c]{@{}l@{}}Regression (Code Coverage)\end{tabular}  & Code Coverage & \begin{tabular}[c]{@{}l@{}}Triangle \\ Classification\end{tabular} \\  \hline
\cite{Shihao22:PredCov} & 2022 & Supervised & Backpropagation NN & \begin{tabular}[c]{@{}l@{}}Existing Inputs, \\ Code Coverage\end{tabular} & \begin{tabular}[c]{@{}l@{}}Regression (Code Coverage)\end{tabular}  & \begin{tabular}[c]{@{}l@{}}Code Coverage, \\ Efficiency\end{tabular} & \begin{tabular}[c]{@{}l@{}}Numeric\\ Functions\end{tabular} \\  \hline
\end{tabular}
}
\caption{
Publications under \textbf{System Test Generation (White Box)} with publication date, ML type, ML technique, training data, objective of the ML, evaluation metrics, and applications used to evaluate. NN = Neural Network.
}
\label{tab:systemTestingWhite}
\end{table}


A total of \paperssystem publications target system testing. Table~\ref{tab:systemTesting} outline Black Box approaches, while Table~\ref{tab:systemTestingWhite} outlines White Box approaches. 
Each table is sorted by ML approach, then by the first author's name. When discussing the objective, we indicate both type of prediction and the purpose of the prediction. 


\smallskip\noindent\textbf{Input Generation (Supervised, Semi-Supervised):} Supervised approaches generally train models that associate particular SUT input with targeted qualities. 
Multiple authors use supervised learning to infer a model from execution logs containing inputs and resulting output~\cite{Bergadano1993a,Budnik2018,Papadopoulos2015,Sharma22:Property}. The model is used to predict input leading to output of interest. For example, Budnik et al. identify small changes in input that lead to large differences in output, indicating boundary areas where faults are likely to emerge~\cite{Budnik2018}. Both Bergadano and Budnik et al. suggest comparing predictions with real output and using misclassifications to indicate the need to re-train~\cite{Bergadano1993a,Budnik2018}. 



Another concern is achieving code coverage. Majma et al. use supervised learning for both input and oracle generation~\cite{Majma2014}. A model associates inputs with paths through the source code, then generates new inputs that execute uncovered paths. 
Similarly, Feldmeier and Fraser generate test inputs for games by targeting code segments, then training neural networks to predict the player actions in particular game states that will cover the associated lines of code~\cite{Feldmeier22:Games}. 
Utting et al. cluster log files---gathered from customer reports---then compare clusters to logs from executing existing test cases to identify weakly-tested areas of the SUT~\cite{Utting2020}. Supervised learning is used to fill in these gaps. These logs are formatted as vectors of actions, and the model predicts the next input in the sequence. 


Others train models to predict which input will fail. Kirac et al. train a model to identify usage behaviors likely to lead to failures using past test cases~\cite{Kirac2019}. 
Eidenbenz et al. randomly generate a set of inputs, execute them, label the execution based on whether they failed, and then cluster failing instances to enhance accuracy~\cite{Eidenbenz2021}. They train a model using several algorithms, then compare their ability to predict failing input. They propose an iterative process where more training data is added over time, and predictions are verified by developers. 


Several authors generate input using models inferred from behavioral specifications (e.g., requirements). The generated input can then show that these specifications are met. Kikuma et al. create a dataset where requirements are tagged with output that should appear if the requirement is met. Their model associates input actions, conditions, and outputs in the requirements, then generates new tests with inputs, conditions, and expected output~\cite{Kikuma2019}.
Ueda et al. transform specifications, written in natural language, into a structured abstract test recipe that can be concretely instantiated with different input~\cite{Ueda2021}. 
Meinke et al. model use cases in a constraint language and generate input from the model inferred from the constraints~\cite{Meinke2021}. 
In addition, both Deng et al. and Zhang et al. generate input for autonomous vehicles intended to violate properties written by human testers~\cite{Deng2021,Zhang18:DeepRoad}. They present adversarial scenarios where multiple neural networks manipulate image data used as input to an autonomous driving system. Collectively, these models predict which input will violate properties by, e.g., changing day to night or adding rain, and use feedback on their actions to retrain. 


Finally, multiple authors generate complex inputs for particular system types. For example, Shrestha~\cite{Shrestha2020a} train a model to generate valid Simulink models---a visual language for modeling and simulation---for testing tool-chains based on the language. 
For compilers, an input is a full program, resulting in a large space of inputs. Zhu et al. restrict the range of inputs to avoid wasted effort~\cite{Zhu2019}. They focus on domain-specific compilers and generate input appropriate for those domains. They extract features from the code, such as number of loops or matrix operations, then train a model to predict whether a new test case belongs to that domain. Test cases not belonging are discarded. 
Protocols require textual input that conforms to a specified format. Often, determining conformance requires manual construction of a grammar. Gao et al. generate protocol test input without a pre-defined grammar~\cite{Gao2019}. Their model learns the probability distributions of every character of a message, enabling the generation of new valid text sequences. 


\smallskip\noindent\textbf{Input Generation (Reinforcement Learning):} Araiza-Illan et al.~\cite{AraizaIllan2016}, Huurman et al.~\cite{Huurman2020}, Shu et al.~\cite{Shu22:EFSM}, and Veanes et al.~\cite{Veanes06:RL} use reinforcement learning to generate input to cover states or transitions of a model. Araiza-Illan et al. generate input for robots~\cite{AraizaIllan2016}. The agent explores the robot's environment, using coverage of plan models as the reward function. 
Huurman et al. model APIs as stateful systems---where requests trigger transitions---and generate API calls intended to cover all transitions in the model~\cite{Huurman2020}. 
Shu et al.~\cite{Shu22:EFSM} and Veanes et al.~\cite{Veanes06:RL} use reinforcement learning to choose input actions for state-based system models. These tests can then be applied to the real system. 

Baumann et al. use reinforcement learning to select input for autonomous driving that violates critical requirements~\cite{Baumann2021}. The reward function encapsulates headway time, time-to-collision, and required longitudinal acceleration to avoid a collision. 
Reddy et al. use reinforcement learning to generate valid complex inputs (e.g., structured documents)~\cite{Reddy2020}. The reward function favors both unique and valid input. As uniqueness depends on previously-generated input, this is not a problem that can easily be solved with supervised learning. 


\smallskip\noindent\textbf{Enhancing Test Generation:} Rather than fully replacing existing test generation methods, ML can also be used to improve their efficiency or effectiveness. A common target for improvement are Genetic Algorithms. A Genetic Algorithm is a search-based method that generates test cases intended to maximize or minimize a fitness function---a domain-specific scoring function, like the reward function in reinforcement learning. 

Buzdalov and Buzdalova use reinforcement learning to modify the fitness function, adding and tuning sub-objectives that assist in optimizing the core objective of the search~\cite{Buzdalov2013a}. 
Zhao and Lv replace the fitness function with a model that predicts which input will cover unseen output behaviors~\cite{Zhao2007}. 
Liu et al., Mishra et al., and Shihao ey al. also replace the fitness function, training a model to predict which code will be covered by input~\cite{Liu22:CovGen,Mishra2011,Shihao22:PredCov}. These models would be used when there is no tool support to measure coverage, or in cases where measuring coverage would be expensive. 

Esnaashari and Damia use reinforcement learning to manipulate tests within the population generated by the Genetic Algorithm by modifying their input~\cite{Esnaashari2021}. 
Paduraru et al. similarly use reinforcement learning to improve the effectiveness of a random testing tool by taking generated input and modifying it to raise its coverage or execution path length~\cite{Paduraru2021}. 
Zhong et al. bias input selection in a fuzzing tool for autonomous driving simulators by learning which seeds for the random number generator are more likely to lead to traffic violations in the simulation~\cite{Zhong22:Fuzz}.
Sharma et al. replace random input generation in property-based testing with a model inferred from system executions~\cite{Sharma22:Property}. The model is submitted, along with properties of interest, to a SMT solver to find input potentially violating the properties. 

Mirabella et al. train a model to predict input validity, allowing a generation framework to filter invalid input before applying it~\cite{Mirabella2021}. 
Luo et al.~\cite{Luo2021} and Chen et al.~\cite{Chen2021i} both enhance constraint solving in symbolic execution. Normally, a fixed timeout is used. Luo et al. instead train a model using multiple methods to predict the time needed to solve a constraint~\cite{Luo2021}. 
Chen et al. use reinforcement learning to identify the optimal solving strategy for a constraint~\cite{Chen2021i}. 

\subsubsection{GUI Test Generation}

\begin{table}[!t]
\resizebox{\columnwidth}{!}{
\begin{tabular}{llllllll}
\hline
\textbf{Ref} & \textbf{Year} &\textbf{ML Approach} & \textbf{Technique} & \textbf{Training Data} & \textbf{ML Objective}  & \textbf{Evaluation Metrics}  & \textbf{Evaluated On} \\ \hline
\cite{Adamo2018}    & 2018     & Reinforcement                  & Q-Learning                  & N/A                    & \begin{tabular}[c]{@{}l@{}}Reward (State Cov.)\end{tabular}  & State Coverage   & F-Droid    \\ \hline
\cite{Ariyurek2021} & 2021 & Reinforcement & \begin{tabular}[c]{@{}l@{}}Monte Carlo Tree Search, Sarsa\end{tabular} & N/A  & \begin{tabular}[c]{@{}l@{}}Reward (Test\\ Goal Coverage)\end{tabular} & Faults Detected & 2D Games  \\ \hline
\cite{Brunetto2021} & 2021 & Reinforcement& Q-Learning   & N/A & \begin{tabular}[c]{@{}l@{}}Reward (State Cov.)\end{tabular}& Qualitative & Resource Planning \\ \hline
\cite{Choi2013a}    & 2013  & Reinforcement                  & Own Technique    & N/A  & \begin{tabular}[c]{@{}l@{}}Reward (State \\Coverage, Loop \\ Interactions)\end{tabular}   & State Coverage               & F-Droid                                               \\ \hline
\cite{Collins2021}  & 2021 & Reinforcement& Deep Q-Network & N/A & \begin{tabular}[c]{@{}l@{}}Reward (State \\ Change Magnitude)\end{tabular}    & \begin{tabular}[c]{@{}l@{}}Code Coverage, \\ Faults Detected \end{tabular}  & F-Droid  \\ \hline
\cite{Degott2019}   & 2019    & Reinforcement                  & Q-Learning                        & N/A                    & \begin{tabular}[c]{@{}l@{}}Reward (State Cov., \\ Element Interaction)\end{tabular} & State Coverage                 & F-Droid                                             \\ \hline
\cite{Khan22:Android} & 2022 & Reinforcement & Sarsa & N/A & \begin{tabular}[c]{@{}l@{}}Reward (Event Cov.)\end{tabular}      & Code Coverage   &  F-Droid   \\ \hline
\cite{Koroglu2020}  & 2020   & Reinforcement                  & Double Q-Learning           & N/A                    & \begin{tabular}[c]{@{}l@{}}Reward (State Cov.,\\ Specifications)\end{tabular}      & State Coverage                 &  F-Droid                                            \\ \hline
\cite{Koroglu2021}  & 2021 & Reinforcement& Double Q-Learning & N/A & \begin{tabular}[c]{@{}l@{}}Reward \\(Specifications)\end{tabular}   & Faults Detected  & F-Droid \\ \hline
\cite{Koroglu2018}  & 2018   & Reinforcement                  & Q-Learning                  & N/A                    & \begin{tabular}[c]{@{}l@{}}Reward (State Cov.,\\ Specifications)\end{tabular}      & State Coverage                     & F-Droid                                          \\ \hline
\cite{Mariani2012}  & 2012  & Reinforcement                  & Q-Learning                  & N/A                    & \begin{tabular}[c]{@{}l@{}}Reward (State Cov., \\ Calls)\end{tabular}      & State Coverage  & \begin{tabular}[c]{@{}l@{}}Password \\ Manager, \\ PDF Reader, \\ Task List, \\ Budgeting\end{tabular}                                         \\ \hline
\cite{Pan2020}      & 2020    & Reinforcement   & Q-Learning + Long Short-Term Memory   & N/A   & \begin{tabular}[c]{@{}l@{}}Reward (State Cov., \\ Curiosity)\end{tabular}  & State Coverage    & \begin{tabular}[c]{@{}l@{}}F-Droid, Other \\ Android Apps\end{tabular}  \\ \hline
\cite{Sherin22:QExplore} & 2022 & Reinforcement & Q-Learning & N/A & \begin{tabular}[c]{@{}l@{}}Reward (Curiosity,\\ Validity of Resulting\\ State)\end{tabular}  & \begin{tabular}[c]{@{}l@{}}Code Coverage,\\ Faults Detected, \\ Input Diversity, \\ State Coverage\end{tabular}    & Web Apps \\ \hline
\cite{Vuong2018}    & 2018 & Reinforcement                  & Q-Learning                  & N/A                    & \begin{tabular}[c]{@{}l@{}}Reward (State Cov.)\end{tabular}    & State Coverage    & Android Apps  \\ \hline
\cite{Yasin2021} & 2021 &  Reinforcement& Q-Learning  & N/A & \begin{tabular}[c]{@{}l@{}}Reward (State Cov.)\end{tabular} & \begin{tabular}[c]{@{}l@{}}Code Coverage, \\ Faults Detected\end{tabular}  & F-Droid  \\ \hline
\cite{Zheng2021a}   & 2021 & Reinforcement&  Q-Learning & N/A & \begin{tabular}[c]{@{}l@{}}Reward (State Cov., \\ Curiosity)\end{tabular} & \begin{tabular}[c]{@{}l@{}}Code Coverage, \\ Faults Detected, \\ Scalability\end{tabular}   & \begin{tabular}[c]{@{}l@{}}Web Apps \\ (Research, \\ Real-World, \\ Industrial)\end{tabular} \\ \hline
\cite{Zheng19:Game} & 2019 & Reinforcement &  Advantage Actor-Critic & N/A &
\begin{tabular}[c]{@{}l@{}}Reward (Game-Specific)\end{tabular} &
\begin{tabular}[c]{@{}l@{}}Faults Detected,\\ State Coverage, \\ Code Coverage\end{tabular}   & \begin{tabular}[c]{@{}l@{}}Games\end{tabular} \\ \hline
\cite{Kamal2019}    & 2019    & Supervised         & Feedforward NN             & Generated Inputs         & Regression (Output) & \begin{tabular}[c]{@{}l@{}}State Coverage\end{tabular}  & Login Web App  \\\hline
\cite{Khaliq22:UI} & 2022 & Supervised & Residual NN, Transformer & \begin{tabular}[c]{@{}l@{}}UI Screenshots,\\ Natural Language\end{tabular} & \begin{tabular}[c]{@{}l@{}}Classification \\ (UI Elements), \\ Regression (Natural\\ Language Test)\end{tabular} & \begin{tabular}[c]{@{}l@{}}Accuracy, \\ Flakiness,\\ Input Validity\end{tabular} & Android Apps \\ \hline
\cite{Khaliq22:Web} & 2022 & Supervised & Residual NN, Transformer & \begin{tabular}[c]{@{}l@{}}UI Screenshots,\\ Natural Language\end{tabular} & \begin{tabular}[c]{@{}l@{}}Classification \\ (UI Elements), \\ Regression (Natural\\ Language Test)\end{tabular} & \begin{tabular}[c]{@{}l@{}}Accuracy, \\ Efficiency,\\ Flakiness,\\ Input Validity\end{tabular} & Web Apps \\ \hline
\cite{Li2019}       &  2019   & Supervised           & Deep NN                     & System Executions       & \begin{tabular}[c]{@{}l@{}}Regression (Action\\ Probability)\end{tabular}      & State Coverage    & Android Apps                       \\ \hline
\cite{Santiago2018} & 2018    & Supervised           & Recurrent NN                & Existing Inputs       & \begin{tabular}[c]{@{}l@{}}Regression \\ (Test Flows)\end{tabular}    & State Coverage   & \begin{tabular}[c]{@{}l@{}}Unspecified\\ Web App\end{tabular}                        \\ \hline
\cite{Santiago2019} & 2019    & Supervised           & Random Forest      & Web Pages & \begin{tabular}[c]{@{}l@{}}Classification \\ (Page Elements)\end{tabular}     & Mutation Score   & \begin{tabular}[c]{@{}l@{}}Task List, \\Job Recruiting\\ Web Apps\end{tabular}                              \\ \hline
\cite{Yazdani21:GUI} & 2021 & Supervised & UNet & Screenshots & \begin{tabular}[c]{@{}l@{}}Regression (Relevant\\ Screen Areas)\end{tabular}  &  \begin{tabular}[c]{@{}l@{}}Adaptivity, \\Code Coverage\end{tabular}  & \begin{tabular}[c]{@{}l@{}}Androtest,\\ Web Apps\end{tabular}  \\ \hline
\end{tabular}
}
\caption{
Publications under \textbf{GUI Test Generation} with publication date, ML type, ML technique, training data, objective of the ML, evaluation metrics, and applications used to evaluate. NN = Neural Network.
}
\label{tab:GUIresults}
\end{table}

Table~\ref{tab:GUIresults} details the \papersgui GUI testing publications. 
GUI test generation often focuses on a state-based interface model that formulates display changes as transitions taken following input. Fifteen publications generated input covering this model. 

Almost all publications adopted reinforcement learning, as it can learn from feedback after applying an action to the GUI, and many GUIs require a sequence of actions to access certain elements. 
The main difference between publications lies in the reward function. Many base the reward on coverage of the states of the interface model (e.g.,~\cite{Yasin2021}), while incorporating additional information to bias state selection. Additional factors include magnitude of the state change~\cite{Collins2021}, usage specifications~\cite{Koroglu2018,Koroglu2020}, unique code functions called~\cite{Mariani2012}, curiosity---favoring exploration of new elements~\cite{Pan2020,Sherin22:QExplore,Zheng2021a}---coverage of interaction methods (e.g. click, drag)~\cite{Degott2019}, validity of the resulting state~\cite{Sherin22:QExplore}, and avoidance of navigation loops~\cite{Choi2013a}. 

Rather than state coverage, Koroglu and Sen base reward on finding violations of specifications~\cite{Koroglu2021}. Ariyurek et al. also apply reinforcement learning to select input for grid-based 2D games~\cite{Ariyurek2021}. The game state is represented as a graph, and ``test goals'' are synthesized from the graph. The reward emphasizes test goal coverage. 
Li et al. use supervised learning, training a model to mimic patterns from interaction logs~\cite{Li2019}. Their model associates GUI elements with a probability of usage---using probabilities to bias action selection. Kamal et al. filter redundant test cases as part of enhancing a search-based test generation framework~\cite{Kamal2019}. Their model associates input and output, then uses the predicted output to decide if tests are redundant. 

Santiago et al. use supervised learning to generate sequences of interactions~\cite{Santiago2018,Santiago2019}. They trained using human-written interaction sequences spanning several web pages. Their second study extends the approach to interact with forms by extracting feedback messages from the forms~\cite{Santiago2019}. The framework learns constraints for form input, and a constraint solver creates input that meets those constraints. A model is used to classify page components. This helps control how different component types are processed. Their approach requires a complex training phase and a large human-created dataset. However, models can be used for multiple websites, decreasing the training burden.

Khaliq et al. employ multiple forms of supervised learning to generate test cases for Android apps~\cite{Khaliq22:UI} and web apps~\cite{Khaliq22:Web} They use an object recognition model to detect UI elements, then use extracted data from the UI elements to prompt a transformer model for a natural language test description. They then use a parser to translate this description into an executable test case. Similarly, Yazdani et al. have trained a model to identify the UI elements relevant to a particular input action, and use the identified elements to focus random test generation for Android apps~\cite{Yazdani21:GUI}. They also demonstrated that the model can be effectively transferred to web apps. 

Zheng et al. use both search-based test generation and a deep reinforcement learning technique to generate input for games~\cite{Zheng19:Game}. Test input is generated by the search algorithm. However, reinforcement learning is used to select policies that control the generation framework.

\subsubsection{Unit Test Generation}

\begin{table}[!t]
\resizebox{\columnwidth}{!}{
\begin{tabular}{lllllllll}
\hline
\textbf{Ref} & \textbf{Year} & \textbf{\begin{tabular}[c]{@{}l@{}}Test Gen. \\ Approach\end{tabular}} & \textbf{ML Approach} & \textbf{Technique}  & \textbf{Training Data}  & \textbf{ML Objective}  & \textbf{Evaluation Metrics} & \textbf{Evaluated On} \\ \hline
\cite{Walkinshaw2017} & 2017  & Black & Supervised  & \begin{tabular}[c]{@{}l@{}}Query Strategy Framework\end{tabular} & System Executions & \begin{tabular}[c]{@{}l@{}}Regression (Output)\end{tabular} & Mutation Score & \begin{tabular}[c]{@{}l@{}}Math Library, \\ Time Library\end{tabular} \\ \hline
\cite{Hooda2018} & 2018 & Black& Unsupervised  & Backpropagation NN & Existing Inputs & \begin{tabular}[c]{@{}l@{}}Clustering \\(Input Similarity)\end{tabular}  & Not Evaluated  & N/A  \\ \hline
\cite{Almulla2020a}   & 2020  & White& Reinforcement  & \begin{tabular}[c]{@{}l@{}}Upper Confidence Bound, \\Differential Semi-Gradient Sarsa\end{tabular} & N/A & \begin{tabular}[c]{@{}l@{}}Reward (Input\\ Diversity)\end{tabular}  & \begin{tabular}[c]{@{}l@{}}Input Diversity, \\ Faults Detected\end{tabular} & \begin{tabular}[c]{@{}l@{}}JSON Parser\end{tabular} \\ \hline
\cite{Almulla2020}    & 2020  & White & Reinforcement  & \begin{tabular}[c]{@{}l@{}}Upper Confidence Bound, \\Differential Semi-Gradient Sarsa\end{tabular} & N/A & \begin{tabular}[c]{@{}l@{}}Reward (Num. \\ Exceptions)\end{tabular}  & \begin{tabular}[c]{@{}l@{}}Num. Exceptions, \\ Faults Detected\end{tabular} & \begin{tabular}[c]{@{}l@{}}Defects4J\end{tabular}\\ \hline
\cite{Almulla22:AFFS} & 2022 & White & Reinforcement  & \begin{tabular}[c]{@{}l@{}}Upper Confidence Bound, \\Differential Semi-Gradient Sarsa\end{tabular} & N/A & \begin{tabular}[c]{@{}l@{}}Reward (Num. \\ Exceptions, Input\\ Diversity, Strong\\Mutation)\end{tabular}  & \begin{tabular}[c]{@{}l@{}}Num. Exceptions, \\ Input Diversity, \\ Mutation Score, \\Faults Detected\end{tabular} & \begin{tabular}[c]{@{}l@{}}Defects4J\end{tabular}\\ \hline
\cite{Groce2011}  & 2011 & White& Reinforcement & Not Specified  & N/A & \begin{tabular}[c]{@{}l@{}}Reward (Code \\ Coverage)\end{tabular}  & Code Coverage & Data Structures \\ \hline
\cite{He2015}  & 2015  & White& Reinforcement  & Q-Learning  & N/A  & \begin{tabular}[c]{@{}l@{}}Reward (Code \\ Coverage)\end{tabular}  & Code Coverage & \begin{tabular}[c]{@{}l@{}}Data Structures, \\ Collection Library, \\ Primitives Library, \\ Java/XML Parsers\end{tabular} \\ \hline
\cite{Kim2018} & 2018  & White& Reinforcement & \begin{tabular}[c]{@{}l@{}}Double Deep Q-Network\end{tabular}  & N/A & \begin{tabular}[c]{@{}l@{}}Reward (Code \\ Coverage)\end{tabular} & \begin{tabular}[c]{@{}l@{}}Code Coverage, \\ Efficiency\end{tabular} & \begin{tabular}[c]{@{}l@{}}GCD, EXP,\\ Remainder\end{tabular} \\ \hline
\cite{Chen22:Baton} & 2022 & White & Supervised & \begin{tabular}[c]{@{}l@{}}Naive Bayes, Random Forest, \\ SVM, J48\end{tabular} & Existing Inputs & \begin{tabular}[c]{@{}l@{}}Classification \\ (Code Coverage)\end{tabular} & \begin{tabular}[c]{@{}l@{}}Accuracy, \\ Code Coverage, \\ Mutation Score,\\ Efficiency\end{tabular}  & \begin{tabular}[c]{@{}l@{}}Numeric Functions, \\Wheel Brake, \\ Rendering, Mine \\Control, Notification, \\XML Parser, \\ Siemens Benchmark\end{tabular}\\ \hline
\cite{Hershkovich2021} & 2021 & White & Supervised  & Gradient Boosting & Code Metrics  & \begin{tabular}[c]{@{}l@{}}Classification \\ (Fault Prediction)\end{tabular} & \begin{tabular}[c]{@{}l@{}}Accuracy, \\ Faults Detected\end{tabular}  & \begin{tabular}[c]{@{}l@{}}Compression, \\Imaging Library, \\Math Library, \\ NLP,  String \\Library\end{tabular}\\ \hline
\cite{Ji2019}  & 2019  & White& Supervised & Backpropagation NN  & Existing Inputs & \begin{tabular}[c]{@{}l@{}}Regression \\ (Code Coverage)\end{tabular}  & Not Evaluated & N/A \\ \hline
\end{tabular}
}
\caption{Publications under \textbf{Unit Test Generation} with publication date, generation approach, ML type, ML technique, training data, objective of the ML, evaluation metrics, and applications used to evaluate. NN = Neural Network.
}
\label{tab:unitTesting}
\end{table}

Because unit testing focuses on individual classes---making domain concerns less applicable---the majority of publications in Table~\ref{tab:unitTesting} are ``White Box'' approaches and are not tied to particular system types. 

Groce~\cite{Groce2011} and Kim et al.~\cite{Kim2018} use reinforcement learning to generate input, with code coverage as the reward. Groce applies reinforcement learning to generate input directly~\cite{Groce2011}. In contrast, Kim et al. use reinforcement learning to generate optimization-based input generation algorithms~\cite{Kim2018}. The agent manipulates heuristics controlling the search algorithms. 

Walkinshaw and Fraser use a supervised approach to generate input for system parts that have only been weakly tested~\cite{Walkinshaw2017}. A model is trained to predict the output. The model will have more confidence in prediction accuracy for input similar to the training data. Input with low certainty is retained, as they are likely to test parts of the system ignored in the training data. These inputs can later be used to re-train the model, shifting focus to other parts of the system. 

Many authors use ML to enhance existing test generation approaches---often based on Genetic Algorithms. Almulla and Gay use reinforcement learning to select which fitness functions will be optimized in service of a higher-level testing goal~\cite{Almulla2020, Almulla2020a, Almulla22:AFFS}. For example, the agent can learn which combinations of fitness functions best trigger exceptions~\cite{Almulla2020}, increase input diversity~\cite{Almulla2020a}, or increase Strong Mutation Coverage~\cite{Almulla22:AFFS}. He et al. use reinforcement learning to improve coverage of private and inherited methods by augmenting generated tests~\cite{He2015}. The agent can make two types of changes---it can replace a method call with one whose return type is a subclass of the original method's, and it can replace a call to a public method with a call to a method that calls a private method. The reward is focused on private method coverage. Chen et al. employ supervised learning to improve the effectiveness of random generation and concolic testing~\cite{Chen22:Baton}. They generate classification models for particular branches in the code, and use predictions about whether a test will cover a branch to ease the constraint-solving process. 

Hershkovich et al. predict whether a class is likely to be faulty~\cite{Hershkovich2021}. This can improve generation efficiency by determining which classes to target. They train a model---using an ensemble of methods---using source code metrics, labeled on whether a class had faults. 
Ji et al. use supervised learning to replace a fitness evaluation in a Genetic Algorithm~\cite{Ji2019}. They focus on data-flow coverage, which is very expensive to calculate. The model replaces the need to actually measure coverage. Hooda et al. train a model to cluster test cases~\cite{Hooda2018}. When new tests are generated, those too close to a cluster centroid are rejected, improving generation efficiency.

\subsubsection{Performance Test Generation}

\begin{table}[!t]
\resizebox{\columnwidth}{!}{
\begin{tabular}{llllllll}
\hline
\textbf{Ref}  & \textbf{Year} & \textbf{ML Approach} & \textbf{Technique} & \textbf{Training Data} & \textbf{ML Objective} & \textbf{Evaluation Metrics} & \textbf{Evaluated On}  \\ \hline
\cite{Ahmad2019}  & 2019 & Reinforcement  & \begin{tabular}[c]{@{}l@{}}Dueling Deep Q-Network\end{tabular} & N/A & \begin{tabular}[c]{@{}l@{}}Reward (Execution Time)\end{tabular} & \begin{tabular}[c]{@{}l@{}}Identified \\ Bottlenecks\end{tabular} & \begin{tabular}[c]{@{}l@{}}Auction Website\end{tabular} \\ \hline
\cite{HelaliMoghadam2019} & 2019 & Reinforcement  & Q-Learning & N/A & \begin{tabular}[c]{@{}l@{}}Reward (Response Time Deviation)\end{tabular} & Not Evaluated   & N/A \\ \hline
\cite{Koo2019}  & 2019 & Reinforcement & Q-Learning & N/A & \begin{tabular}[c]{@{}l@{}}Reward (Path Length, \\Feasibility)\end{tabular} & \begin{tabular}[c]{@{}l@{}}Paths Explored, \\ Efficiency\end{tabular} & \begin{tabular}[c]{@{}l@{}} Biological Computation,\\ Parser, Sorting, \\ Data Structures\end{tabular} \\ \hline
\cite{Moghadam2019} & 2019 & Reinforcement  & Q-Learning & N/A & \begin{tabular}[c]{@{}l@{}}Reward (Response Time Deviation)\end{tabular} & Not Evaluated  & N/A   \\ \hline
\cite{Moghadam22:PerfTest} & 2022 & Reinforcement & Q-Learning + Fuzzy Logic & N/A & \begin{tabular}[c]{@{}l@{}}Reward (Response Time Deviation,\\ Resource Usage)\end{tabular} & \begin{tabular}[c]{@{}l@{}}Efficiency, \\Adaptivity \end{tabular}  & \begin{tabular}[c]{@{}l@{}}Resource-Sensitive \\Programs (e.g., \\ compression)\end{tabular}\\ \hline
\cite{Chen22:DeepPerform} & 2022 & Semi-Supervised & \begin{tabular}[c]{@{}l@{}}Generative Adversarial Network\end{tabular}& System Executions & \begin{tabular}[c]{@{}l@{}}Regression (Performance), \\ Classification (Input Distribution)\end{tabular} & \begin{tabular}[c]{@{}l@{}} Identified Bottlenecks, \\ Efficiency, \\ State Coverage, \\ Model Sensitivity, \\ Input Quality\end{tabular}& Object Recognition \\ \hline
\cite{Sedaghatbaf2021} & 2021 & Semi-Supervised  & \begin{tabular}[c]{@{}l@{}}Conditional Generative \\Adversarial Network\end{tabular} & System Executions & \begin{tabular}[c]{@{}l@{}}Regression (Perf. Requirements), \\ Classification (Test Realism)\end{tabular} & \begin{tabular}[c]{@{}l@{}} Identified Bottlenecks, \\ Accuracy, Labelling \\and Training Effort\end{tabular}& Auction Website \\ \hline
\cite{Luo2016} & 2016 & Supervised & RIPPER & System Executions & \begin{tabular}[c]{@{}l@{}} Regression (Rule Learning)\end{tabular} & \begin{tabular}[c]{@{}l@{}}Identified  Bottlenecks\end{tabular} & \begin{tabular}[c]{@{}l@{}}Insurance, Online Stores, \\Project Management\end{tabular} \\ \hline
\cite{Schulz2021} & 2021 & Supervised  & \begin{tabular}[c]{@{}l@{}}Multivariate Time Series\end{tabular} & Session Logs & \begin{tabular}[c]{@{}l@{}}Regression (Load)\end{tabular} & Accuracy & Student Information \\ \hline
\end{tabular}
}
\caption{
Publications under \textbf{Performance Test Generation} with publication date, ML type, ML technique, training data, objective of the ML, evaluation metrics, and applications used to evaluate. NN = Neural Network.
}
\label{tab:performance}
\end{table}

Performance test generation refers to the generation of test cases for the purpose of assessing whether the SUT meets non-functional requirements, such as speed, response time, scalability, or resource usage requirements~\cite{Ahmad2019}. Such tests are often generated to identify and eliminate performance bottlenecks. Table~\ref{tab:performance} details the performance testing publications. Performance can be measured, which offers feedback for subsequent rounds of generation. Thus, the majority of approaches are based on iterative processes, including reinforcement~\cite{Ahmad2019,Koo2019,Moghadam2019,HelaliMoghadam2019}, rule~\cite{Luo2016}, and adversarial learning~\cite{Sedaghatbaf2021}. 

Ahmad et al. generate input intended to expose performance bottlenecks, with reward based on maximized execution time~\cite{Ahmad2019}. 
They note room for improvement by integrating other performance indicators into the reward.
Rather than generating explicit program input, Moghadam et al. apply reinforcement learning to control the execution environment~\cite{Moghadam2019,HelaliMoghadam2019,Moghadam22:PerfTest}. They identify resource configurations (CPU, memory, disk) where timing requirements are violated, with reward based on response time deviation and resource usage. These environmental factors constitute ``implicit'' input that can change the behavior of the SUT. 

Sedaghatbaf et al. generate input violating performance requirements using two competing neural networks~\cite{Sedaghatbaf2021}. The generator produces input, and the discriminator classifies whether input violates requirements. This feedback improves the generator. 
Chen et al. also employ adversarial learning to generate input for resource-constrained neural networks intended to expose performance bottlenecks for such networks~\cite{Chen22:DeepPerform}. 

Luo et al. use the RIPPER rule learner to identify input classes that trigger intensive computations~\cite{Luo2016}. When tests are executed, executions are clustered based on execution time. RIPPER learns and iteratively refines rules differentiating the clusters, which are then used to generate new input. 

Schulz et al. generate workloads for load testing~\cite{Schulz2021}. The model generates realistic load levels on a system at various times and scenarios. Past session logs are clustered, and a multivariate time series is applied to predict system load during a scenario. 
Finally, Koo et al. use reinforcement learning to improve symbolic execution during stress testing~\cite{Koo2019}. They identify input that triggers worst-case execution time, defined as inputs that trigger a long execution path. The agent controls the exploration policy used by symbolic execution so that long paths are favored. The reward is based on path length and the feasibility of generating input for that path. 


\subsubsection{Combinatorial Interaction Testing}

\begin{table}[t]
\resizebox{\columnwidth}{!}{
\begin{tabular}{llllllll}
\hline
Ref  & Year & ML Approach  & Technique  & Training Data & ML Objective  & Evaluation Metrics & Evaluated On    \\ \hline
\cite{Jia2015}   & 2015 & Reinforcement  & SOFTMAX & N/A  & \begin{tabular}[c]{@{}l@{}}Reward (Input \\ Combinations)\end{tabular} & \begin{tabular}[c]{@{}l@{}}Covering Array Size, \\ Efficiency\end{tabular}  & \begin{tabular}[c]{@{}l@{}}Misc. Synthetic, \\ Real Systems\end{tabular} \\ \hline
\cite{Mudarakola2018} & 2018 & Supervised  & Artificial NN & Specifications & \begin{tabular}[c]{@{}l@{}}Other (Structure\\ Input Space), \\ Regression (Output)\end{tabular} & Covering Array Size & \begin{tabular}[c]{@{}l@{}}Temperature \\Monitoring\end{tabular} \\ \hline
\cite{Mudarakola2014} & 2014 & Supervised  & Artificial NN  & \begin{tabular}[c]{@{}l@{}}Pairwise Input \\ Combinations\end{tabular} &  \begin{tabular}[c]{@{}l@{}}Other (Structure\\ Input Space)\end{tabular} & Covering Array Size & Web Apps  \\ \hline
\cite{Patil2018}      & 2018 & Supervised   & Artificial NN  & \begin{tabular}[c]{@{}l@{}}Pairwise Input \\ Combinations\end{tabular} & \begin{tabular}[c]{@{}l@{}}Regression (Input \\ Coverage)\end{tabular} & \begin{tabular}[c]{@{}l@{}}Covering Array Size, \\ Efficiency\end{tabular} & Unspecified  \\ \hline
\cite{DuyNguyen2013}  & 2013 & Unsupervised & \begin{tabular}[c]{@{}l@{}}Expectation-Maximization\end{tabular} & System Executions & \begin{tabular}[c]{@{}l@{}}Clustering (Code\\ Coverage)\end{tabular} & Qualitative Analysis & \begin{tabular}[c]{@{}l@{}}Bubble Sort, Math \\Functions, HTTP \\Processing, Banking \end{tabular}\\ \hline
\end{tabular}
}
\caption{
Publications under \textbf{Combinatorial Interaction Testing} with publication date, ML type, ML technique, training data, objective of the ML, evaluation metrics, and applications used to evaluate. NN = Neural Network.
}
\label{tab:cit}
\end{table}

Table~\ref{tab:cit} shows the publications that use ML as part of Combinatorial Interaction Testing (CIT). Mudarakola et al.~\cite{Mudarakola2014,Mudarakola2018} and Patil and Prakash~\cite{Patil2018} use neural networks to generate covering arrays---minimal sets of tests that cover all pairwise interactions between input variables. Patil and Prakash~\cite{Patil2018} predict the interactions covered by an input. They use this model to identify a covering array. Mudarakola et al~\cite{Mudarakola2014} map each hidden layer of a neural network to a variable and each node to a value class. The values are connected by their connection to other variables. They do not use the network for prediction, but as a structuring mechanism to generate a covering array. Code coverage is used to prune redundant test cases. In a follow-up study~\cite{Mudarakola2018}, they manually construct a network using requirements, linking outputs to input values, with each input node mapping to an input variable, hidden layers linked to conditions from the requirements, and output nodes linked to predicted SUT output. The network again provides structure---a covering array is generated based on paths through the network. 


Jia et al. use reinforcement learning to tune the generation strategy of a search-based generation framework using Simulated Annealing~\cite{Jia2015}. The agent selects how Simulated Annealing mutates a covering array. The reward is based on the change in coverage of combinations after imposing a mutation. Their framework recognizes and exploits policies that improve coverage. 

CIT assumes that input values are divided into classes. Division is generally done manually, but identifying divisions is non-trivial. Duy Nguyen and Tonella use clustering to identify value classes, based on executed code lines (and how many times lines were executed)~\cite{DuyNguyen2013}. 


\subsubsection{Test Oracle Generation}\label{sec:oracles}

\begin{table}[!t]
\resizebox{\columnwidth}{!}{
\begin{tabular}{llllllll}
\hline
\textbf{Ref}  & \textbf{Year} & \textbf{ML Approach} & \textbf{Technique} & \textbf{Training Data} & \textbf{ML Objective}  & \textbf{Evaluation Metric} & \textbf{Evaluated On} \\ \hline
\cite{Braga2018}    & 2018  & Supervised & Adaptive Boosting & System Executions & \begin{tabular}[c]{@{}l@{}}Classification (Verdict)\end{tabular}  & Mutation Score  & Shopping Cart \\ \hline
\cite{Chen2021a}   & 2021  & Supervised   & Convolutional NN &  Screenshots & \begin{tabular}[c]{@{}l@{}}Classification (Verdict)\end{tabular}  & \begin{tabular}[c]{@{}l@{}}Accuracy, \\ Faults Detected\end{tabular} & Games (Android, iOS)\\ \hline
\cite{Gholami2018} & 2018  & Supervised & Backpropagation NN & System Executions & \begin{tabular}[c]{@{}l@{}}Classification (Verdict)\end{tabular}  & Mutation Score & Embedded Software  \\ \hline
\cite{Ibrahimzada22:Oracle} & 2022 & Supervised & Recurrent NN & \begin{tabular}[c]{@{}l@{}}Source/Test Code,\\ System Executions\end{tabular} & Classification (Verdict) & \begin{tabular}[c]{@{}l@{}}Accuracy,\\ Efficiency, \\ Faults Detected,\\ Mutation Score\end{tabular} &  Defects4J \\ \hline
\cite{Kamaraj22:oracle} & 2022 & Supervised & Artificial NN & System Executions & Classification (Verdict) & Correct Classifications & Not Specified \\ \hline
\cite{Khosrowjerdi2018} & 2018  & Supervised   & L*  & System Executions      & \begin{tabular}[c]{@{}l@{}}Classification (Verdict)\end{tabular}    & \begin{tabular}[c]{@{}l@{}}Faults Detected, \\ Efficiency\end{tabular}  & Platoon Simulator \\ \hline
\cite{Khosrowjerdi2017} & 2017  & Supervised   & Not Specified & System Executions      & \begin{tabular}[c]{@{}l@{}}Classification (Verdict)\end{tabular}    & Faults Detected    & Automotive Applications        \\ \hline
\cite{Makondo2016}   & 2016  & Supervised  & Multilayer Perceptron & System Executions  & \begin{tabular}[c]{@{}l@{}}Classification (Verdict)\end{tabular}  & Accuracy & User Creation\\ \hline
\cite{Rafi22:PredART} & 2022 & Supervised & \begin{tabular}[c]{@{}l@{}}Convolutional NN, \\Multilayer Perceptron\end{tabular} & Screenshots & \begin{tabular}[c]{@{}l@{}}Regression(Deviation\\ from Correctness)\end{tabular} & Accuracy & Augmented Reality Apps\\ \hline
\cite{Shahamiri2010a} & 2010 & Supervised   & Backpropagation NN   & System Executions      & \begin{tabular}[c]{@{}l@{}}Classification (Verdict)\end{tabular}   & Mutation Score    & Student Registration        \\ \hline
\cite{Tsimpourlas2021}    & 2021   & Supervised & \begin{tabular}[c]{@{}l@{}}Multilayer Perceptron, \\Long Short-Term Memory NN\end{tabular}  & System Executions &  \begin{tabular}[c]{@{}l@{}}Classification (Verdict)\end{tabular}   & \begin{tabular}[c]{@{}l@{}}Accuracy, Training \\ Data Size\end{tabular} & \begin{tabular}[c]{@{}l@{}}Blockchain Module, \\ Deep Learning Module, \\ Encryption Library, \\ Stream Editor\end{tabular} \\ \hline
\cite{Tsimpourlas22:Oracle}    & 2022   & Supervised & \begin{tabular}[c]{@{}l@{}}Multilayer Perceptron,  \\Long Short-Term Memory NN\end{tabular}  & System Executions &  \begin{tabular}[c]{@{}l@{}}Classification (Verdict)\end{tabular}   & \begin{tabular}[c]{@{}l@{}}Accuracy, Adaptivity, \\Training Data Size\end{tabular} & \begin{tabular}[c]{@{}l@{}}Blockchain Module, \\ Deep Learning Module, \\ Encryption Library, \\ Network Protocols,\\ Stream Editor,\\ String Library\end{tabular} \\ \hline
\end{tabular}
}
\caption{Publications under \textbf{Test Verdicts Test Oracle Generation} with publication date, ML type, ML technique, training data, objective of the ML, evaluation metrics, and applications used to evaluate. NN = Neural Network.}
\label{tab:oracle}
\end{table}

\begin{table}[!t]
\resizebox{\columnwidth}{!}{
\begin{tabular}{llllllll}
\hline
\textbf{Ref}  & \textbf{Year} & \textbf{ML Approach} & \textbf{Technique} & \textbf{Training Data} & \textbf{ML Objective}  & \textbf{Evaluation Metric} & \textbf{Evaluated On} \\ \hline
\cite{Aggarwal2004}  & 2004  & Supervised & Backpropagation NN    & System Executions    & \begin{tabular}[c]{@{}l@{}}Classification (Output)\end{tabular}  & Correct Classifications & Triangle Classification \\ \hline
\cite{Arrieta2021}   & 2021  & Supervised   & \begin{tabular}[c]{@{}l@{}}Regression Tree, Support Vector Machine, \\Ensemble, RGP, Stepwise Regression\end{tabular} & System Executions & \begin{tabular}[c]{@{}l@{}}Regression (Time)\end{tabular} & Accuracy & Elevator \\ \hline
\cite{Dinella2022:TOGA} & 2022 & Supervised & Transformer & Source/Test Code & Regression (Assertions) & 
\begin{tabular}[c]{@{}l@{}}Faults Detected, \\Accuracy\\ \end{tabular}
& \begin{tabular}[c]{@{}l@{}}Defects4J\end{tabular} \\ \hline
\cite{Ding2016} & 2016 & Supervised & Support Vector Machine  & System Executions      & \begin{tabular}[c]{@{}l@{}}Classification (Output)\end{tabular}   & Mutation Score  & Image Processing   \\ \hline
\cite{Gartziandia2021} & 2021 & Supervised  & \begin{tabular}[c]{@{}l@{}}Regression Tree, Support Vector Machine, \\Ensemble, TRGP, Stepwise Regression\end{tabular} & System Executions & \begin{tabular}[c]{@{}l@{}}Regression (Time)\end{tabular} & Accuracy & Elevator\\ \hline
\cite{Gartziandia22:elevator} & 2022 & Supervised & \begin{tabular}[c]{@{}l@{}}Regression Tree, Support Vector Machine, \\Ensemble, RGP, Stepwise Regression\end{tabular} & System Executions & \begin{tabular}[c]{@{}l@{}}Regression (Waiting,\\ Execution Time)\end{tabular} & Mutation Score, Accuracy & Elevator\\ \hline
\cite{Jin2008}  &2008   & Supervised & Backpropagation NN    & System Executions & \begin{tabular}[c]{@{}l@{}}Classification (Output)\end{tabular}   & Correct Classifications  & Triangle Classification \\ \hline
\cite{Majma2014} & 2014 & Supervised  & Backpropagation NN & System Executions & \begin{tabular}[c]{@{}l@{}}Classification (Output)\end{tabular}  & Faults Detected & Static Analysis\\  \hline 
\cite{Monsefi2019}  &2019   & Supervised & Deep NN   & System Executions & \begin{tabular}[c]{@{}l@{}}Regression (Output)\end{tabular} & Mutation Score   & Mathematical Functions  \\ \hline
\cite{Sangwan2011}   &2011 & Supervised  & Radial-Basis Function NN & System Executions  & \begin{tabular}[c]{@{}l@{}}Regression (Output)\end{tabular}  & Correct Classifications & Triangle Classification \\ \hline
\cite{Shahamiri2011} &2011  & Supervised   & Multilayer Perceptron & System Executions  & \begin{tabular}[c]{@{}l@{}}Classification (Output)\end{tabular}    & Mutation Score    & Insurance Application   \\ \hline
\cite{Shahamiri2012} & 2012  & Supervised   & Multilayer Perceptron & System Executions     & \begin{tabular}[c]{@{}l@{}}Classification (Output)\end{tabular}  & Mutation Score       & Insurance Application   \\ \hline
\cite{Shahamiri10:Oracle} & 2010  & Supervised   & Artificial NN & System Executions  & \begin{tabular}[c]{@{}l@{}}Classification (Output)\end{tabular}    & \begin{tabular}[c]{@{}l@{}}Mutation Score, Accuracy,\\ Precision, Correct \\ Classifications\end{tabular}    & Student Registration   \\ \hline
\cite{Singhal2016}  & 2016   & Supervised   & \begin{tabular}[c]{@{}l@{}}Backpropagation NN + Cascade\end{tabular} & System Executions      & \begin{tabular}[c]{@{}l@{}}Classification (Output)\end{tabular}  & Accuracy & Credit Analysis     \\ \hline
\cite{Vanmali2002}   & 2002  & Supervised   & Not Specified  & System Executions      & \begin{tabular}[c]{@{}l@{}}Classification (Output)\end{tabular}    & Mutation Score     & Credit Analysis   \\ \hline
\cite{Vineeta2014a} & 2014  & Supervised  & \begin{tabular}[c]{@{}l@{}}Backpropagation NN, Decision Tree\end{tabular} & System Executions      & \begin{tabular}[c]{@{}l@{}}Classification (Output)\end{tabular}     & Mutation Score         & Triangle Classification    \\ \hline
\cite{Ye06:Oracle} & 2006  & Supervised   & Backpropagation NN & System Executions      & \begin{tabular}[c]{@{}l@{}}Regression (Output)\end{tabular}   & Precision &  Mathematical Functions  \\ \hline
\cite{Ye2006a}    & 2006  & Supervised   & Multilayer Perceptron & System Executions      & \begin{tabular}[c]{@{}l@{}}Regression (Output)\end{tabular}   & Mutation Score &  Mathematical Functions  \\ \hline
\cite{Yu22:ATLAS} & 2022 & Supervised & Transformer & Source/Test Code & Regression (Assertions) & 
Accuracy & \begin{tabular}[c]{@{}l@{}}Misc. Open-Source \\Projects\end{tabular} \\ \hline
\cite{Zhang2019d}   & 2019  & Supervised   & Probabilistic NN & System Executions      & \begin{tabular}[c]{@{}l@{}}Classification (Output)\end{tabular}                       & Correct Classifications & Prime, Triangle Class \\ \hline
\end{tabular}
}
\caption{Publications under \textbf{Expected Output Test Oracle Generation} with publication date, ML type, ML technique, training data, objective of the ML, evaluation metrics, and applications used to evaluate. NN = Neural Network.}
\label{tab:oracle2}
\end{table}

\begin{table}[!t]
\resizebox{\columnwidth}{!}{
\begin{tabular}{llllllll}
\hline
\textbf{Ref}  & \textbf{Year} & \textbf{ML Approach} & \textbf{Technique} & \textbf{Training Data} & \textbf{ML Objective}  & \textbf{Evaluation Metric} & \textbf{Evaluated On} \\ \hline
\cite{Hiremath2020}  & 2020 & Reinforcement & Not Specified  & N/A & \begin{tabular}[c]{@{}l@{}}Reward (Relations)\end{tabular} & Not Evaluated & \begin{tabular}[c]{@{}l@{}}Ocean Modeling\end{tabular}    \\ \hline
\cite{JHiremath2021}  & 2021 & Reinforcement & Not Specified  & N/A & \begin{tabular}[c]{@{}l@{}}Reward (Relations)\end{tabular}& Not Evaluated & \begin{tabular}[c]{@{}l@{}}Ocean Modeling\end{tabular}     \\ \hline
 \cite{Spieker2020} & 2020 & Reinforcement & Contextual Bandit & N/A  & \begin{tabular}[c]{@{}l@{}}Reward (Faults Detected)\end{tabular} & Faults Detected & \begin{tabular}[c]{@{}l@{}}Object  Detection\end{tabular} \\ \hline
\cite{Hardin2018}   &2018  & Supervised & Support Vector Machine  & Code Features & \begin{tabular}[c]{@{}l@{}}Classification (Property)\end{tabular} & Accuracy &  \begin{tabular}[c]{@{}l@{}}Misc. Functions\end{tabular} \\ \hline
\cite{Kanewala2013b} &2013 & Supervised   & \begin{tabular}[c]{@{}l@{}}Support Vector Machine, Decision Trees\end{tabular}   & Code Features & \begin{tabular}[c]{@{}l@{}}Classification (Property)\end{tabular} & Mutation Score & \begin{tabular}[c]{@{}l@{}}Misc. Functions\end{tabular} \\ \hline
\cite{Kanewala2016}  & 2016  & Supervised    & Support Vector Machine   & Code Features  & \begin{tabular}[c]{@{}l@{}}Classification (Property)\end{tabular} & Mutation Score  & \begin{tabular}[c]{@{}l@{}}Misc. Functions\end{tabular}\\ \hline
\cite{Korkmaz2021}   & 2021 & Supervised   & Decision Trees & System Executions  & \begin{tabular}[c]{@{}l@{}}Regression (Conditions)\end{tabular} & Accuracy & \begin{tabular}[c]{@{}l@{}}Android Apps\end{tabular} \\ \hline
\cite{Nair2019}  & 2019  & Supervised   & Support Vector Machine      & Code Features    &  \begin{tabular}[c]{@{}l@{}}Classification (Property)\end{tabular} & ROC & \begin{tabular}[c]{@{}l@{}}Matrix Calculation\end{tabular}   \\ \hline
\cite{Shu2007}    & 2007  & Supervised  & L*  & System Executions & \begin{tabular}[c]{@{}l@{}}Classification (Violation)\end{tabular} & Training Data Size & \begin{tabular}[c]{@{}l@{}}Handshake Protocols\end{tabular}   \\ \hline
\cite{Zhang2017}    & 2017  & Supervised  & Radial-Basis Function NN & Code Features     & \begin{tabular}[c]{@{}l@{}}Classification (Property)\end{tabular} & Accuracy & \begin{tabular}[c]{@{}l@{}}Misc. Functions\end{tabular}\\ \hline
\end{tabular}
}
\caption{Publications under \textbf{Metamorphic (And Other Properties) Test Oracle Generation} with publication date, ML type, ML technique, training data, objective of the ML, evaluation metrics, and applications used to evaluate. NN = Neural Network.}
\label{tab:oracle3}
\end{table}

Tables~\ref{tab:oracle}-\ref{tab:oracle3} summarize the \papersoracle oracle generation publications. Almost all approaches adopt supervised learning. 
These approaches train models, which stand in for traditional oracles, using previous system executions, screenshots, or metadata about source code features. The model predicts the correctness of output or properties of the expected output.

\smallskip\noindent\textbf{Test Verdicts:} The majority of studies employ neural networks to train models that directly predict whether a test should pass or fail~\cite{Ibrahimzada22:Oracle,Kamaraj22:oracle,Makondo2016,Shahamiri2010a,Gholami2018,Tsimpourlas2021,Tsimpourlas22:Oracle}. Most are simple, traditional neural networks for simple programs. However, Ibrahimzada et al. and Tsimpourlas et al. have recently explored how deep learning can train models for complex programs~\cite{Ibrahimzada22:Oracle,Tsimpourlas2021,Tsimpourlas22:Oracle}. Braga et al. also are able to generate models for a complex application using an ensemble technique~\cite{Braga2018}.


Chen et al. train a model to identify rendering errors in video games by training on screenshots of previous faults~\cite{Chen2021a}. 
Rafi et al. apply a similar process to identify object-placement errors in augmented reality apps~\cite{Rafi22:PredART}. They do not predict a concrete pass/fail verdict, as users may perceive object placement differently. Instead, when labelling data, they asked multiple humans to offer verdicts, then labeled examples with the percentage that responded with a pass verdict. The model, then, predicts the percentage of users that would see a placement as correct in a new screenshot.

Khosrowjerdi et al. combine supervised learning and model checking~\cite{Khosrowjerdi2017}. A model is learned from system executions that predicts output. Given the model and specifications, a model checker assesses whether each specification is met, yielding a verdict. For each violation, a test is generated that can be executed to confirm the fault. If the fault is not real, the test and its outcome can be used to retrain the model. 
In a follow-up study~\cite{Khosrowjerdi2018}, they demonstrate their technique on systems-of-systems.

\smallskip\noindent\textbf{Expected Output:} The approaches generally train on system executions, and then predict the specific output expected for a new input. Output is often abstracted to representative values or limited to functions with enumerated values, rather than specific output. For example, a common application is ``triangle classification''---a classification of a triangle as scalene, isosceles, equilateral, or not-a-triangle. This function is often used as an initial demonstration for test generation algorithms because it has branching behavior. Because it has limited outputs, it is also a common target for demonstrating the potential of oracle generation. Zhang et al. model a function that judges whether an integer is prime---a binary classification problem\cite{Zhang2019d}. Many others also generate oracles for applications with a limited range of output~\cite{Ding2016,Shahamiri10:Oracle,Shahamiri2011,Shahamiri2012,Singhal2016,Vanmali2002}. However, some authors have generated oracles for functions with unconstrained---e.g., integer---output~\cite{Arrieta2021,Gartziandia2021,Gartziandia22:elevator,Monsefi2019,Ye2006a,Ye06:Oracle}.  

The majority of approaches used some form of neural network~\cite{Aggarwal2004, Jin2008, Majma2014,Monsefi2019,Sangwan2011,Shahamiri10:Oracle,Shahamiri2011, Shahamiri2012,Singhal2016,Vineeta2014a,Ye2006a,Ye06:Oracle,Zhang2019d}. Ding and Zhang~\cite{Ding2016} also used label propagation---a technique where labeled and unlabeled training data are used, and the algorithm propagates labels to similar, unlabeled data---to reduce the quantity of labeling to create the training data. 

Recently, Dinella et al.~\cite{Dinella2022:TOGA} and Yu et al.~\cite{Yu22:ATLAS} demonstrated the use of language-generating transformer models for test oracle creation. Rather than inferring a model from system executions, a model is trained instead on source and test code, then given the code-under test and/or a partial unit test, the model directly produces assertions predicted to be appropriate for the prompt. Such models are trained on large datasets of code from many projects, and can potentially be applied generally.


\smallskip\noindent\textbf{Metamorphic and Properties:} Several publications build on the research of Kanewala and Bieman~\cite{Kanewala2013b}, whose approach (a) converts code into control-flow graphs, (b) selects code elements as features for a data set, and (c), trains a model that predicts whether a feature exhibits a particular metamorphic relation from a list. This requires training data where features are labeled with a classification based on whether or not they exhibit a particular relation. 
Kanewala et al. extended this work by adding a graph kernel~\cite{Kanewala2016}. Hardin and Kanawala adapted this approach for label propagation~\cite{Hardin2018} . Zhang et al. extended the approach to a multi-label classification that can handle multiple metamorphic relations at once~\cite{Zhang2017}. 
Finally, Nair et al. demonstrated how data augmentation can enlarge the training dataset using mutants as the source of additional training data~\cite{Nair2019}. 

Korkmaz and Yilmaz predict the conditions on screen transitions in a GUI~\cite{Korkmaz2021}. Their model is trained using past system execution and potential guard conditions. 
Shu and Lee use supervised learning to assess security properties of protocols~\cite{Shu2007}. A protocol is specified using a state machine, and message confidentiality is assessed on message reachability. A model is inferred, then assessed for violations. If a violation is found, input is produced to check against the implementation. If the violation is false, the test helps retrain the model.  

Hiremath et al. predict metamorphic relations for ocean modeling~\cite{Hiremath2020, JHiremath2021}. The reinforcement learning approach poses relations, evaluates whether they hold, and attempts to minimize a cost function based on the validity of the set of proposed relations. 
Spieker and Gotlieb use reinforcement learning to \textit{select} metamorphic relations from a superset of potentially-applicable relations~\cite{Spieker2020}. Their approach evaluates whether selected relations can discover faults in an image classification algorithm. 


\subsection{RQ2: Goals of Applying ML}\label{sec:rq2}

Table~\ref{tab:rq2} lists the goals of authors in adopting ML, sorted into three broad categories. In the first two, ML is used directly to generate input or an oracle. As previously discussed, oracle generation uses ML to predict output, to properties of output, or a test verdict. Regarding input generation, the most common goal is to use ML to increase coverage of some criterion associated with effective testing. This includes coverage of code, states or transitions of models, or input interactions. Other uses of ML include generating input that exposes performance bottlenecks, demonstrates conformance to---or violation of---specifications, or increases input/output diversity. Others generate input for a complex data type or input likely to fail. 

\begin{table}[!t]
\centering
\resizebox{\columnwidth}{!}{
\begin{tabular}{llrl}
\hline
\textbf{Type of Goal} & \textbf{Goal} & \textbf{\# Pubs.} & \textbf{Publications}\\ \hline
\multirow{7}{*}{\textbf{Generate Input}} & Maximize Coverage & 32 & \begin{tabular}[c]{@{}l@{}}\cite{Adamo2018,AraizaIllan2016,Ariyurek2021,Brunetto2021,Choi2013a,Collins2021,Degott2019,Feldmeier22:Games,Groce2011,Huurman2020,Khaliq22:UI,Khaliq22:Web}\\
\cite{Khan22:Android,Kim2018,Koroglu2020,Koroglu2018,Li2019,Majma2014,Mariani2012,Mudarakola2018,Mudarakola2014,Pan2020,Patil2018} \\
\cite{Santiago2018,Santiago2019,Sherin22:QExplore,Shu22:EFSM,Utting2020,Veanes06:RL,Vuong2018,Yasin2021,Zheng2021a}
\end{tabular} \\ \cline{2-4}
& Expose Performance Bottlenecks & 8 & \cite{Ahmad2019,Chen22:DeepPerform,HelaliMoghadam2019,Luo2016,Moghadam2019,Moghadam22:PerfTest,Schulz2021,Sedaghatbaf2021} \\ \cline{2-4}
& Show Conformance to (or Violation of) Specifications & 7 & \cite{Baumann2021,Deng2021,Kikuma2019,Koroglu2021,Meinke2021,Ueda2021,Zhang18:DeepRoad}\\ \cline{2-4}
& Generate Complex Inputs & 4 & \cite{Gao2019,Reddy2020,Shrestha2020a,Zhu2019} \\ \cline{2-4}
& Improve Input or Output Diversity & 4 & \cite{Bergadano1993a,Budnik2018,DuyNguyen2013,Walkinshaw2017}\\ \cline{2-4}
& Predict Failing Input & 2 & \cite{Eidenbenz2021,Kirac2019} \\ \hline
\multirow{3}{*}{\textbf{Generate Oracle}} & Predict Output & 20 & 
\begin{tabular}[c]{@{}l@{}}
\cite{Aggarwal2004,Arrieta2021,Dinella2022:TOGA,Ding2016,Gartziandia2021,Gartziandia22:elevator,Jin2008,Majma2014,Monsefi2019,Sangwan2011,Shahamiri2011,Shahamiri2012} \\
\cite{Shahamiri10:Oracle,Singhal2016,Vanmali2002,Vineeta2014a,Ye06:Oracle,Ye2006a,Yu22:ATLAS,Zhang2019d}
\end{tabular} \\  \cline{2-4}
& Predict Test Verdict & 12 & \cite{Braga2018,Chen2021a,Gholami2018,Ibrahimzada22:Oracle,Kamaraj22:oracle,Khosrowjerdi2018,Khosrowjerdi2017,Makondo2016,Rafi22:PredART,Shahamiri2010a,Tsimpourlas2021,Tsimpourlas22:Oracle} \\  \cline{2-4}
& Predict Properties of Output & 10 & \cite{Hardin2018,Hiremath2020,JHiremath2021,Kanewala2013b,Kanewala2016,Korkmaz2021,Nair2019,Shu2007,Spieker2020,Zhang2017}\\ \hline
\multirow{2}{*}{\textbf{Enhance Existing Method}} & Improve Effectiveness & 15 & \begin{tabular}[c]{@{}l@{}} \cite{Almulla2020a,Almulla2020,Almulla22:AFFS,Buzdalov2013a,Chen22:Baton,Esnaashari2021,He2015,Jia2015,Koo2019,Paduraru2021,Sharma22:Property,Yazdani21:GUI}\\
\cite{Zhao2007,Zheng19:Game,Zhong22:Fuzz}
\end{tabular}\\ \cline{2-4}
& Improve Efficiency & 10 & \cite{Chen2021i,Hershkovich2021,Hooda2018,Ji2019,Kamal2019,Liu22:CovGen,Luo2021,Mirabella2021,Mishra2011,Shihao22:PredCov}\\ 
\hline
\end{tabular}}
\caption{ML goals and the number of publications pursuing each goal.}
\label{tab:rq2}
\end{table}

In the final category, ML tunes the performance or effectiveness of a generation framework---often search-based of Symbolic Execution-based approaches.  To improve efficiency, ML clusters redundant tests, replaces expensive calculations with predictions, chooses generation targets, or checks input validity. To improve effectiveness, ML manipulates test cases (e.g., replaces method calls) or tunes the generation strategy (e.g., selects fitness functions, mutation heuristics, or timeouts).
 
\begin{center}
\begin{framed}
  \textbf{RQ2 (Goal of ML):} ML generates input (47\%)---particularly to maximize some form of coverage---or oracles (33\%)---particularly that predict an expected output. It is also used to improve efficiency or effectiveness of existing generation methods (20\%). 
\end{framed}
\end{center}

\subsection{RQ3: Integration into Test Generation}


RQ3 highlights where and how ML has been integrated into the testing process. This includes types of ML applied, training data, and how ML was used (regression, classification, reward functions). 

Supervised techniques were the first applied to input and oracle generation, and remain the most common. Supervised techniques are---by far---the most common for oracle generation. They are also the most common for system and combinatorial interaction testing. The predictions made by models are either from pre-determined options (classification) or open (regression). Classification is often used in oracle generation, e.g., to produce a verdict (pass/fail) or output from a limited range. Regression is common in input generation, where complex predictions must be made. 

Both training time and quantity of training data need to be accounted for when considering a supervised technique. After being trained, a model will not learn from new interactions, unlike with reinforcement learning. A model must be retrained with new training data to improve its accuracy. Therefore, it is important that supervised methods be supplied with sufficient quantity and quality of training data. Supervised techniques generally learn from past system executions, labeled with a measurement of interest. If the label can be automatically recorded, then gathering sufficient data is often not a major concern. However, if the SUT is computationally inefficient or information is not easily collectible (e.g., a human must label data), it can be difficult to use supervised ML. 

Adversarial learning may help overcome data challenges. This strategy forces models to compete, creating a feedback loop where performance is improved without the need for human input. Multiple publications adopted adversarial networks, generally in cases where input was associated with a numeric quality (performance, vehicle speed---e.g.,~\cite{Deng2021,Zhang18:DeepRoad}). Neither case requires human labeling, so models can be automatically retrained. Other recent deep learning approaches---often trained on many systems---show promise in their ability to adapt to unseen systems (e.g.,~\cite{Tsimpourlas22:Oracle,Zhong22:Fuzz}).

Reinforcement learning is the second most common type of ML. Reinforcement learning was even used more often than supervised in 2020, and almost as often in 2021. Reinforcement learning has been used in all input generation problems and is the most common technique for GUI, unit, and performance generation. 

Reinforcement learning is appealing because it does not require pre-training and automatically improves accuracy through interactions. Reinforcement learning is most applicable when effectiveness can be judged using a numeric metric, i.e., where a measurable assessment already exists. This includes performance measurements---e.g., resource usage---or code coverage. Reinforcement learning is also effective when the SUT has branching or stateful behavior---e.g., in GUI testing, where a \textit{sequence} of input may be required. Similarly, performance bottlenecks often emerge as the consequence of a sequence of actions, and code coverage may require multiple setup steps. Reinforcement learning is effective in such situations because it can learn from the outcome of taking an action. Therefore, it is effective at constructing sequences of input steps that ultimately achieve some goal of interest. Many supervised approaches are not equipped to learn from each individual action, and must attempt to predict the full sequence of steps at once. 

Outside of individual tests, reinforcement learning is also effective at enhancing test generation algorithms. Genetic Algorithms, for example, evolve test suites over a series of subsequent generations. Reinforcement learning can tune aspects of this evolution, in some cases guided by feedback from the same fitness functions targeted by the optimization. If a test suite attains high fitness, reinforcement learning may be able to improve that score by manipulating the test cases of the algorithm parameters. Reinforcement learning can, of course, generate input effectively in a similar manner to an optimization algorithm. However, it also can often improve the algorithm such that it produces even better tests. 

Authors of sampled publications applied unsupervised learning to cluster test cases to improve generation efficiency or to identify weakly tested areas of the SUT. While clustering has not been used often in the sampled publications, clustering is common in other testing practices (e.g., to identify tests to execute~\cite{surveyMLinTesting2019}). Therefore, it may have potential for use in filtering tasks during generation, especially to improve efficiency. Future work should further consider how clustering could be applied as part of test generation.

\begin{center}
\begin{framed}
   \textbf{RQ3 (Integration of ML):} The most common ML types are supervised (61\%) and reinforcement learning (34\%). Some publications also employ unsupervised (2\%) or semi-supervised (3\%) learning. (Semi-)Supervised learning is the most common ML for system testing, CIT, and all forms of oracle. Reinforcement learning is the most common technique for GUI, unit, and performance testing, and is used where testing goals often have measurable scores, a sequence of input is required, or existing generation tools can be tuned. Clustering was also used for filtering, e.g., discarding similar test cases.
\end{framed}
\end{center}

\subsection{RQ4: ML Techniques Applied}

\begin{table}[!t]
\begin{center}
\resizebox{\columnwidth}{!}{
\begin{tabular}{lllr}
\hline
\textbf{Type}  & \textbf{Family} & \textbf{Technique} & \textbf{\# Pubs.} \\ \hline
\multirow{16}{*}{\textbf{Supervised}}  & \multirow{8}{*}{\textbf{Neural Networks}} & Backpropagation NN & 14   \\ \cline{3-4} 
& & Multi-Layer Perceptron & 8 \\ \cline {3-4}
& & Artificial NN & 7 \\ \cline {3-4}
& & Long Short-Term Memory NN & 6 \\ \cline {3-4}
& & Transformer & 4 \\ \cline{3-4}
& & Radial-Basis Function NN & 3 \\ \cline{3-4}
& & Convolutional NN, Deep NN, Feedforward NN, Recurrent NN, Residual NN & 2 \\ \cline {3-4}
& & \begin{tabular}[c]{@{}l@{}}Backpropagation NN + Cascade, Probabilistic NN, Shallow NN, UNet \end{tabular} & 1 \\ \cline{2-4}
& \multirow{4}{*}{\textbf{Trees}} & Decision Tree & 5 \\ \cline{3-4} 
& & Random Forest & 3 \\ \cline{3-4}
& & Gradient Boosting, Regression Tree & 2 \\ \cline{3-4}
& & \begin{tabular}[c]{@{}l@{}}Ada-Boosted Tree, C4.5, J48, Tree-LSTM\end{tabular} & 1 \\ \cline{2-4}
& \multirow{5}{*}{\textbf{Others}} & Support Vector Machine & 11 \\ \cline{3-4} 
& & \begin{tabular}[c]{@{}l@{}}L*, Conditional Random Fields, Ensemble, K-Nearest Neighbors, \\ Regression Gaussian Process, Stepwise Regression\end{tabular} & 2 \\ \cline{3-4}
& & \begin{tabular}[c]{@{}l@{}}Adaptive Boosting, Gaussian Process, Multivariate Time Series, \\Naive Bayes, Parallel Distributed Processing, Query Strategy Framework, RIPPER\end{tabular} & 1 \\ \hline  
\multirow{11}{*}{\textbf{Reinforcement}} & \multirow{6}{*}{\textbf{Q-Learning}}& Q-Learning & 16 \\ \cline{3-4} 
& & Deep Q-Network & 3 \\ \cline{3-4}
& & Double Q-Learning & 2 \\ \cline{3-4}
& & \begin{tabular}[c]{@{}l@{}}Delayed Q-Learning, Dueling Deep Q-Network, Double Deep Q-Network,\\ Q-Learning + Fuzzy Logic, Q-Learning + Long Short-Term Memory, \\ReLU Q-Learning\end{tabular} & 1 \\ \cline{2-4}
& \multirow{4}{*}{\textbf{Others}} & \begin{tabular}[c]{@{}l@{}}Differential Semi-Gradient Sarsa, Upper Confidence Bound\end{tabular} & 3 \\ \cline{3-4}  
& & Sarsa & 2 \\ \cline{3-4}
& & \begin{tabular}[c]{@{}l@{}}Advantage Actor-Critic, Asynchronous Advantage Actor Critic, Contextual Bandit, \\  
Markov Decision Process,  Monte Carlo Control, Monte Carlo Tree Search, SOFTMAX
\end{tabular} & 1 \\ \hline 
\multirow{2}{*}{\textbf{Semi-Supervised}}  & & Generative Adversarial Network & 3 \\ \cline{3-4}
& & \begin{tabular}[c]{@{}l@{}}Convolutional NN, Conditional Generative Adversarial Network\end{tabular} & 1 \\ \hline
\textbf{Unsupervised}   & &  Backpropagation NN, Expectation-Maximization, MeanShift & 1   \\ \hline
\end{tabular}
}\caption{ML techniques adopted---divided by ML type and family of ML techniques---ordered by number of publications where the technique is adopted. NN = Neural Network.}
\label{tab:techniquesoverall}
\end{center}
\end{table}

RQ4 examines specific ML techniques. Table~\ref{tab:techniquesoverall} lists techniques employed, divided by ML type. Neural networks are the most common techniques in supervised learning. Support vector machines are also employed often, as are forms of decision trees. 

In particular, backpropagation neural networks are used most (11\%). Backpropagation neural networks are a classic technique where a network is composed of multiple layers~\cite{HECHTNIELSEN199265}. In each layer, a weight value for each node is calculated. In such networks, information is fed \textit{forward}---there are no cyclic connections to earlier layers. However, the backpropagation feature propagates error backward, allowing earlier nodes to adjust weights if necessary. This leads to less complexity and faster learning rates. In recent years, more complex neural networks have continued to implement backpropagation as one (of many) features. 

Recently, neural networks utilizing Long Short-Term Memory have also become quite common. Unlike traditional feedforward neural networks, Long Short-Term Memory has feedback connections~\cite{Graves2012}. This creates loops in the network, allowing information to persist. This adaptation allows such networks to process not just single data points, but sequences where one data point depends on earlier points. Long Short-Term Memory networks and deep neural networks are likely to become more common in the next few years as more researchers adopt deep learning techniques. 

The emergence of transformer models---complex neural networks that learn from, and generate, natural language~\cite{Dinella2022:TOGA}---is promising for both test and oracle generation. Transformers make use of a mechanism called ``self-attention'' that uses backpropagation to infer the relationship between words in a phrase~\cite{NIPS2017_3f5ee243}. This mechanism enables automated context-extraction and summarization of text, which in term enables the model to produce complex textual output as well. 

Reinforcement learning is dominated by forms of Q-Learning---Q-Learning and its variants are used in 22\% of publications. Q-Learning is a prototypical form of off-policy reinforcement learning, meaning that it can choose either to take an action guided by the current ``best'' policy---maximizing expected reward---or it can choose to take a random action to refine the policy~\cite{Sutton2018}. Many other reinforcement learning techniques are also off-policy, and follow a similar process, with various differences (e.g., calculating reward or action decisions in a different manner). 

Some authors have chosen specific techniques because they worked well in previous work (e.g.,~\cite{Koroglu2020,Kanewala2016}). Others saw certain techniques work on similar problems outside of test generation (e.g.,~\cite{Jia2015}), or chose techniques thought to represent the state-of-the-art for a problem class (e.g.,~\cite{Kirac2019}). However, most authors do not justify their choice of technique, nor do they often compare alternatives.

In recent years, open-source ML frameworks have emerged that accelerate the pace and effectiveness of research by making robust algorithms available. The authors of 51 publications (41\% of the sample) explicitly made use of existing frameworks. The most common ML frameworks used in the sampled publications include keras-rl (e.g,~\cite{Kim2018}), Matlab (e.g.,~\cite{Arrieta2021}), OpenAI Gym (e.g.,~\cite{Huurman2020}), PyTorch (e.g.,~\cite{Spieker2020}), scikit-learn (e.g.,~\cite{Hardin2018}), TensorFlow (e.g.,~\cite{Budnik2018}), and WEKA (e.g.,~\cite{Luo2016}). In the other 73 publications---especially older publications---authors either implemented ML algorithms or adapted unspecified implementations. The use of a framework constrains technique choice. However, all of these frameworks offer many techniques, and may allow researchers to compare results across techniques. This could lead to more informed and robust implementations. 

\begin{center}
\begin{framed}
  \textbf{RQ4 (ML Techniques):} Neural networks, especially backpropagation neural networks, are the most common supervised techniques. Reinforcement learning is generally based on Q-Learning. Technique choice is often not explained, but may be inspired by insights from previous or related work, an algorithm having performed well on a similar problem, or algorithms available in open-source frameworks (e.g., OpenAI Gym or WEKA).  
\end{framed}
\end{center}

\subsection{RQ5: Evaluation of the Test Generation Framework}

\begin{table}[!t]
\begin{center}
\begin{tabular}{llr}
\hline
\textbf{Type}     & \textbf{Metric}  & \textbf{\# Pubs.} \\ \hline
\multirow{8}{*}{\textbf{Supervised}}  & \begin{tabular}[c]{@{}l@{}} \textbf{Prediction Accuracy (e.g., correct classifications, ROC)}\end{tabular}   & 37  \\ \cline{2-3} 
& \begin{tabular}[c]{@{}l@{}}Faults Detected (including mutants and performance issues)\end{tabular}   & 33  \\ \cline{2-3} 
& \begin{tabular}[c]{@{}l@{}}Efficiency (e.g., scalability, \# tests generated/executed, time), \\ Coverage Attained (e.g, code, state)\end{tabular}  & 12 \\ \cline{2-3}
& Test Size (e.g., size of test cases, suite, or covering array) & 4 \\ \cline{2-3}
& \begin{tabular}[c]{@{}l@{}}\textbf{Adaptivity (whether a model can be transferred to a new system)}, \\ Validity of Generated Inputs, \textbf{Quantity of Training Data Required}\end{tabular} & 3 \\ \cline{2-3}
& Flakiness of Generated Tests & 2 \\ \cline{2-3}
& \begin{tabular}[c]{@{}l@{}} Input/Output Diversity, \textbf{Model Size}, \textbf{Sensitivity of Predictions} \end{tabular}  & 1 \\ \hline
\multirow{7}{*}{\textbf{Reinforcement}} & Coverage & 25  \\ \cline{2-3} 
& Faults Detected & 13  \\ \cline{2-3} 
& Efficiency  & 6   \\ \cline{2-3} 
& Input/Output Diversity & 4 \\ \cline{2-3}
& \# Exceptions Discovered & 2 \\ \cline{2-3}
& \begin{tabular}[c]{@{}l@{}}\textbf{Adaptivity}, Qualitative Analysis, \# Queries Solved, \\\# Requirements Met, \textbf{Sensitivity}, Test Size\end{tabular} & 1    \\  \hline
\multirow{3}{*}{\textbf{Semi-Supervised}} & Faults Detected  & 4 \\ \cline{2-3}
& \begin{tabular}[c]{@{}l@{}}\textbf{Prediction Accuracy}, Coverage, Efficiency, Quality of Generated Inputs, \\Validity of Generated Inputs, \textbf{Required Labeling and Training Effort}, \\\textbf{Sensitivity of Predictions}\end{tabular} & 1   \\ \hline
\textbf{Unsupervised} & \textbf{\# Clusters Produced}, Qualitative Analysis  & 1      \\  \hline
\end{tabular}
\caption{Evaluation metrics adopted (similar metrics are grouped), divided by ML approach, and ordered by number of publications using each metric. Metrics in bold are related to ML.}
\label{tab:evaluationsOverall}
\end{center}
\end{table}

RQ5 examines how authors have evaluated their work---in particular, how ML affects evaluation.  The metrics adopted by the authors are listed in Table~\ref{tab:evaluationsOverall}. We group similar metrics (e.g., coverage metrics, notions of fault detection, etc.). In most cases, these metrics are used to evaluate the quality of the input or oracle generation approach. 

In most cases, the entire framework is evaluated. Almost all of these evaluations employ standard metrics for test generation. Some metrics are specific to a testing practice (e.g., covering array size) or aspect of generation (e.g., number of queries solved), while others are applied across testing practices (e.g., fault detection). Naturally---whether ML is incorporated or not---a generation framework must be evaluated on its effectiveness. 

Many authors also evaluate the ML component separately. Supervised approaches were often evaluated using some notion of model accuracy---using various accuracy measurements, correct classification rate, and ROC. Approaches have also been evaluated on the quantity of required training data, whether a model can be applied to unknown systems, and the sensitivity of model predictions to small changes in the input or model parameters. In addition, one study used the size of the trained model to help explain the results of applying the technique, rather than using it to measure solution quality. Semi-supervised approaches were also evaluated using accuracy, the required labeling/training effort, and sensitivity. Finally, one study employing an unsupervised approach used the number of clusters produced to analyze the results of applying their approach. 

Reinforcement learning approaches were generally not evaluated using ML-specific metrics, except for a study that examined their adaptivity and sensitivity. This is reasonable, as reinforcement learning learns how to maximize a numeric function. The reward is based on the goals of the overall generation framework. Rather than evaluating using an absolute notion of accuracy, the success of reinforcement learning can be seen in improved reward measurements, attainment of a checklist of goals, or metrics such as fault detection.

\begin{center}
\begin{framed}
  \textbf{RQ5 (Evaluation):} The full generation framework is generally evaluated by traditional testing metrics (e.g., fault detection). However, the ML components are also evaluated---especially in supervised learning---using accuracy, adaptivity, quantity of training data needed, labeling/training effort, prediction sensitivity, and other ML metrics. Reinforcement learning is generally evaluated using testing metrics tied to the reward.  
\end{framed}
\end{center}

\subsection{RQ6: Limitations and Open Challenges}\label{sec:rq6}

The sampled publications show great potential. However, we have observed multiple challenges that must be overcome to transition research into real-world use.

\smallskip\noindent\textbf{Volume, Contents, and Collection of Training Data:} (Semi-)Supervised ML requires training data to create a model. There are multiple challenges related to the \textit{required volume} of training data, the \textit{required contents} of the training data, and \textit{human effort} required to produce that training data.

Regardless of the testing practice addressed, the volume of required training data can be vast. This data is generally attained from labeled execution logs, which means that the SUT needs to be executed \textit{many} times to gather the information needed to train the model. Approaches based on deep learning could produce highly accurate models but may require thousands of executions to gather required training data. Some approaches also must preprocess the collected data. While it may be possible to automatically gather training data, the time required to produce the dataset can still be high and must be considered. 

This is particularly true for cases where a regression is performed rather than a classification---e.g., an expected value oracle~\cite{Arrieta2021} or complex test input~\cite{Shrestha2020a}. Producing a complex continuous value is more difficult than a simple classification, and requires significant training data---with a range of outcomes---to make accurate predictions.

In addition, the contents of the training data must be considered. If generating input, the training data must contain a wide range of input scenarios with diverse outcomes that reflect the specific problem of interest and its different branching possibilities. Consider code coverage prediction (e.g.,~\cite{Majma2014,Ji2019}). If one wishes to predict the input that will cover a particular element, then the training data must contain sufficient information content to describe how to cover that element. That requires a diverse training set. 

Models based on output behavior---e.g., expected value oracles or models that predict input based on particular output values~\cite{Bergadano1993a,Budnik2018,Papadopoulos2015}---suffer from a related issue. The training data for expected value oracles must either come from passing test cases---that is, the output must be correct---or labels must be applied by humans. A small number of cases accidentally based on failing output may be acceptable if the algorithm is resilient to noise in the training data, but training on faulty code can result in an inaccurate model. This introduces a significant barrier to automating training by, e.g., generating input and simply recording the output that results.

Similarly, models that make predictions based on failures---e.g., test verdict oracles or models that produce input predicted to trigger a failure~\cite{Kirac2019} or performance issue~\cite{Luo2016}---require training data that contains a large number of \textit{failing test cases}. This implies that faults have already been discovered and, presumably, fixed before the model is trained. This introduces a paradox. There may be remaining failures to discover. However, the more training data that is needed, the less the need for---or impact of---the model.

In some cases, training data must be labeled (or even collected) by a human. Again, oracles suffer heavily from this problem. Test verdict oracles require training data where each entry is assigned a verdict. This requires either existing test oracles---reducing the need for a ML-based oracle---or human labeling of test results. Judging test results is time-consuming and can be erroneous as testers become fatigued~\cite{testOracleSurvey2014}, making it difficult to produce a significant volume of training data. Generation of metamorphic relation oracles requires overcoming a similar dilemma, where training data must be labeled based on whether a particular metamorphic relation holds. This requires labeling by a tester with significant knowledge of the source code. 

For some problems, these issues can be avoided by employing reinforcement learning instead. Reinforcement learning will learn while interacting with the SUT. In cases where the effectiveness of ML can be measured automatically---e.g., code coverage and performance bottlenecks---reinforcement learning is a viable solution. However, cases where ground truth is required---e.g., oracles---are not as amenable to reinforcement learning. Reinforcement learning also requires many executions of the SUT, which can be an issue if the SUT is computationally expensive or otherwise difficult to execute and monitor, such as when specialized hardware is required for execution.

Otherwise, techniques are required that (1) can enhance training data, (2) can extrapolate from limited training data, and (3), can tolerate noise in the training data. Means of generating synthetic training data, like in the work of Nair et al.~\cite{Nair2019}, demonstrate the potential for data augmentation to help overcome this limitation. Adversarial learning also offers a way to improve the accuracy of a model---reducing the need for a large training dataset. Again, however, such approaches are of limited use in cases where human involvement is required. In addition, deep learning approaches---such as transformers---can often be trained on data from many different projects, potentially yielding models that are also effective on projects not in their training set (e.g.,~\cite{Yu22:ATLAS}). 

\begin{center}
\begin{framed}
  \textbf{RQ6 (Challenges):} Supervised learning is limited by the required quantity, quality, and contents of training data---especially when human effort is required.  Oracles particularly suffer from these issues. Reinforcement learning and adversarial learning are viable alternatives when data collection and labeling can be automated.
\end{framed}
\end{center}

\smallskip\noindent\textbf{Retraining and Feedback:} After training, models have a fixed error rate and do not learn from new mistakes made. If the training data is insufficient or inaccurate, the generated model will be inaccurate. The ability to improve the model based on additional feedback could help account for limitations in the initial training data. 

There are two primary means to overcome this limitation---either retraining the model using an enriched training dataset or adopting a reinforcement learning approach that can adapt its expectations based on feedback. Both means carry challenges. Retraining requires (a) establishing a schedule for when to train the updated model, and (b), an active effort on the part of human testers to enrich and curate the training dataset. Adversarial learning offers an automated means to retrain the model. However, there are still limitations on when it can be applied.

Enriching the dataset---as well as the use of reinforcement learning---requires some kind of feedback mechanism to judge the effectiveness of the predictions made. This can be difficult in some cases, such as test oracles, where human feedback may be required. Human feedback, even on a subset of the decisions made, reduces the cost savings of automation. 

\begin{center}
\begin{framed}
  \textbf{RQ6 (Challenges):} Models should be retrained over time. How often retraining occurs depends, partially, on the cost to gather and label additional data or on the amount of human feedback required.
\end{framed}
\end{center}

\smallskip\noindent\textbf{Complexity of Studied Systems:} Regardless of ML type, many of the proposed approaches are evaluated on highly simplistic systems. 44\% of the publications evaluate using toy examples, with only a few lines of code or possible function outcomes. While it is intuitive to \textit{start} with simplistic examples to examine the viability of an ML approach, the real-world application requires accurate predictions for complex functions and systems with many branching code paths. If a function is simple, there is likely little need for a predictive model in the first place. Several recent studies feature thorough evaluations of complex systems (e.g.,~\cite{Chen2021i,Almulla22:AFFS,Zheng2021a}), even on industrial systems (e.g.,~\cite{Mirabella2021,Jia2015}). However, many studies evaluate on only a single example or a handful of examples, and many of those examples are still not very complex. It largely remains to be seen whether many proposed techniques can be used on real-world production code. 

The generation of models for arbitrary systems with unconstrained output may be prohibitively difficult even for sophisticated ML techniques. This is particularly the case for expected value oracles. In such cases, some abstraction should be expected---either a simplification of the core logic of the system or a partition of inputs or outputs into symbolic values. One possibility to consider is a variable level of abstraction---e.g., a training-time decision to cluster output predictions into an adjustable number of representative values (such as the centroids of clusters of outputs). Training could take place over different settings for this parameter, and the balance between accuracy and abstraction could be explored.   

In any evaluation, a variety of systems should be considered. The complexity of the systems should vary. This enables the assessment of scalability of the proposed techniques. Researchers should examine how prediction accuracy, training data requirements (for supervised learning), and time to convergence on an optimal policy (for reinforcement learning) scale as the complexity of the system increases. This would enable a better understanding of the limitations and applicability of ML-based techniques in test generation for real-world systems. 

\begin{center}
\begin{framed}
  \textbf{RQ6 (Challenges):} Scalability of ML techniques to real-world systems is not clear. When modeling complex functions, varying degrees of abstraction could be explored if techniques are unable to scale. In evaluations, a range of systems should be considered, and explicit analyses of scalability (e.g., accuracy, training, learning rate) should be performed. 
\end{framed}
\end{center}

\smallskip\noindent\textbf{Variety, Complexity, and Tuning of ML Techniques:} Authors rarely explain or justify their choice of ML algorithm---often stating that an algorithm worked well previously or that it is ``state-of-the-art'', if any rationale is offered. It is even rarer that multiple algorithms are compared to determine which is best for a particular task. As the purpose of many research studies is to demonstrate the viability of an idea, the choice of algorithm is not always critically important. However, this choice still has implications, as it may give a false impression of the applicability of an approach and unnecessarily introduce a performance ceiling that could be overcome through the consideration of alternative techniques.

One reason for this limitation may be that testing researchers are generally ML \textit{users}, not ML experts. They may lack the expertise to know which algorithms to apply. Collaboration with ML researchers may help overcome this challenge. The use of open-source ML frameworks can also ease this challenge by removing the need for researchers to develop their own algorithms. Rather than needing to understand each algorithm, they could instead compare the performance of available alternatives. This comparison would also lead to a richer evaluation and discussion. 

Many of the proposed approaches---especially earlier ones---are based on simple neural networks with few layers. These techniques have strict limitations in the complexity of the data they can model and have been replaced by more sophisticated techniques. Deep learning, which may utilize many hidden layers, may be essential in making accurate predictions for complex systems. Few approaches to date have utilized deep learning, but such approaches are starting to appear, and we would expect more to explore these techniques in the coming years. However, deep learning also introduces steep requirements on the training data that may limit its applicability. 

Almost all of the proposed approaches utilize a single ML technique. An approach explored in many domains is an \textit{ensemble}~\cite{Eidenbenz2021}. In such approaches, models are trained on the same data using a variety of techniques. Each model is asked for a prediction, and then the final prediction is based on the consensus of the ensemble. Ensembles are often able to reach stable, accurate conclusions in situations where a single model may be inaccurate. A small number of studies have applied ensembles~\cite{Eidenbenz2021,Hershkovich2021,Braga2018,Arrieta2021,Gartziandia2021}, but such techniques are rare. 

Many ML techniques have parameters that can be tuned (e.g., learning rate, number of hidden units, or activation function). Parameter tuning can significantly impact prediction accuracy and enable significant improvements in the results of even simple ML techniques. The sampled publications do not explore the impact of such tuning. This is an oversight that should be corrected in future work.

\begin{center}
\begin{framed}
  \textbf{RQ6 (Challenges):}  Researchers rarely justify the choice of ML technique or compare alternatives. The use of open-source ML frameworks can ease comparison. Deep learning and ensemble techniques, as well as hyperparameter tuning, should also be explored more widely.
\end{framed}
\end{center}

\smallskip\noindent\textbf{Lack of Standard Benchmarks:} Research benchmarks have enabled sophisticated analyses and comparison of approaches for automated test generation. Such benchmarks usually contain a set of systems prepared for a particular type of evaluation. Bug benchmarks, in particular, contain real faults curated from a variety of systems, along with metadata on those faults. Such benchmarks ease comparison with past research, remove bias from system selection and demonstrate the effectiveness of techniques. Only a small subset of the sampled publications make use of existing research benchmarks. The most common, by far, is the F-Droid Android benchmark (e.g.,~\cite{Collins2021,Degott2019}). Others made use of examples commonly used in research such as the Defects4J (e.g.,~\cite{Almulla22:AFFS,Dinella2022:TOGA}) or the RUBiS web app example (e.g.,~\cite{Sedaghatbaf2021}). However, the majority of studies do not use benchmarks or open-source evaluation targets. 

Some studies require their own particular evaluation. However, in cases where evaluation is over-simplistic, or where code or metadata is unavailable, this makes comparison and replication difficult. Benchmarks are typically tied to particular system types or testing practices. In cases where benchmarks exist---unit, web app, mobile app, and performance testing in particular---we would encourage researchers to use these benchmarks to enable comparison to past work or to allow researchers to make comparisons with their work. 

In other cases, the creation of benchmarks specifically for ML-enhanced test generation research could advance the state-of-the-art in the field, spur new research advances, and enable replication and extension of proposed approaches. In particular, we recommend the creation of such a benchmark for oracle generation. Such a benchmark should contain a variety of code examples from multiple domains and of varying levels of complexity. Code examples should be paired with the metadata needed to support oracle generation. This would include sample test cases and human-created test oracles, at minimum. Such a benchmark could also include sample training data that could be augmented over time by researchers. 

\smallskip\noindent\textbf{Lack of Replication Package or Open Code:} A common dilemma is lack of access to research code and data. Often, a publication is not sufficient to allow replication or application in a new context. This applies to research in ML-enhanced test generation as well, as only 33\% of the publications in our sample provided open-source code or replication packages. 

Outside of this 33\%, some publications made use of open-source ML frameworks. This is positive, in that the specific ML techniques are trustworthy and available. Potentially, experimental results could be replicated in such cases by applying the same techniques to the same settings. However, there still may not be enough information in the paper to enable replication, such as specific parameter settings. Further, these frameworks evolve over time, and the results may differ because the underlying ML technique has changed since the original study was published. 

Researchers should include a replication package with their source code, execution scripts, and the versions of external dependencies used when the study was performed. This package should also include training data and the gathered experiment observations used by the authors in their analyses. 

\begin{center}
\begin{framed}
\textbf{RQ6 (Challenges):} Research is limited by the overuse of simplistic examples, the lack of common benchmarks, and the unavailability of code and data. Researchers should be encouraged to use available benchmarks, and provide replication packages and open code. New benchmarks could be created for ML challenges (e.g., oracle generation).
\end{framed}
\end{center}

\section{Threats to Validity}\label{sec:threats}

\noindent\textbf{External and Internal Validity:} Our conclusions are based on the publications sampled. It is possible that we may have omitted important publications. This can affect internal validity---the evidence we use to make conclusions---and external validity---the generalizability of our findings. Secondary studies can be valuable even if they do not capture all publications from a research field as long as their selection protocol (search string, inclusion/exclusion criteria, snowballing) ensures an adequate sample to infer similar findings to a complete set of relevant publications. We believe that our selection strategy was appropriate. We tested different search strings and performed a validation exercise to test the robustness of our string. We have used four databases, covering the majority of relevant venues, and performed additional snowballing. Our final set of publications includes \papers primary publications, which we believe is sufficient to make informed conclusions. 

\smallskip\noindent\textbf{Conclusion Validity:} Subjective judgments are part of article selection, data extraction, and categorizing publications. To control for bias, protocols were discussed and agreed upon by both authors, and independent verification took place on---at least---a sample of all decisions made by either author. 

\smallskip\noindent\textbf{Construct Validity:} We used a set of properties to guide data extraction. These properties may have been incomplete or misleading. However, we have tried to establish properties that were informed by our research questions. These properties were iteratively refined, and we believe they have allowed us to thoroughly answer the questions. 

\section{Conclusions} \label{sec:conclusions}

Automated test generation is a well-studied research topic, but there are critical limitations to overcome. Recently, researchers have begun to use ML to enhance automated test generation. We have characterized emerging research on this topic through a systematic mapping study examining testing practices that have been addressed, the goals of using ML, how ML is integrated into the generation process, which specific ML techniques are applied, how the full test generation process is evaluated, and open research challenges. 

We observed that ML generates input for system, GUI, unit, performance, and combinatorial testing or improves the performance of existing generation methods. ML is also used to generate test verdicts, property-based, and expected output oracles. Supervised learning---often based on neural networks---and reinforcement learning---often based on Q-learning---are common, and some publications also employ unsupervised or semi-supervised learning. (Semi-/Un-)Supervised approaches are evaluated using both traditional testing metrics and ML-related metrics (e.g., accuracy), while reinforcement learning is often evaluated using testing metrics tied to the reward function. 

The work-to-date shows great promise, but there are open challenges regarding training data, retraining, scalability, evaluation complexity, ML algorithms employed---and how they are applied---benchmarks, and replicability. Our findings can serve as a roadmap for both researchers and practitioners interested in the use of ML as part of test generation.

\section{Acknowledgments}

This research was supported by Vetenskapsr{\aa}det grant 2019-05275.

\bibliography{refs}

\end{document}